\documentclass[preprints,article,accept,moreauthors,pdftex]{Definitions/mdpi} 

\usepackage{xspace}
\usepackage{bm}
\usepackage{Definitions/aas_macros}
\usepackage{amsmath,amssymb}
\usepackage{upgreek}
\usepackage{braket}

\firstpage{1} 
\pubvolume{1}
\issuenum{1}
\articlenumber{0}
\pubyear{2021}
\copyrightyear{2020}
\datereceived{} 
\dateaccepted{} 
\datepublished{} 
\hreflink{https://doi.org/} 

\Title{Landau Levels in a Gravitational Field: The Schwarzschild Spacetime Case}

\TitleCitation{Landau Levels in a Gravitational Field: The Schwarzschild Spacetime Case}

\Author{Alexandre Landry 
 $^{1}$\orcidA{} and Fay\c{c}al Hammad $^{1,2,3,}$*\orcidB{}}

\AuthorNames{Alexandre Landry and Fay\c{c}al Hammad}

\AuthorCitation{Landry, A.; Hammad, F.}

\address{%
$^{1}$ \quad D\'epartement de Physique, Universit\'e de Montr\'eal,
2900 Boulevard \'Edouard-Montpetit,
\mbox{Montr\'eal, QC H3T 1J4,
Canada;} alexandre.landry.1@umontreal.ca\\
$^{2}$ \quad Department of Physics and Astronomy, Bishop's University, 2600 College Street, \mbox{Sherbrooke, QC J1M 1Z7, Canada}\\
$^{3}$ \quad Physics Department, Champlain 
College-Lennoxville, 2580 College Street, \mbox{Sherbrooke,  
QC J1M~0C8, Canada}}

\corres{Correspondence: fhammad@ubishops.ca}

\abstract{We investigate the gravitational effect on Landau levels. We show that the familiar infinite Landau degeneracy of the energy levels of a quantum particle moving inside a uniform and constant magnetic field is removed by the interaction of the particle with a gravitational field. Two independent approaches are used to solve the relevant Schr\"odinger equation within the Newtonian approximation. It is found that both approaches yield qualitatively similar results within their respective approximations. With the goal of clarifying some results found in the literature concerning the use of a third independent approach for extracting the quantization condition based on a similar differential equation, we show that such an approach cannot yield a general and yet consistent result. We point out to the more accurate, but impractical, way to use such an approach; a way which does in principle yield a consistent quantization condition. We discuss how our results could be used to contribute in a novel way to the existing methods for testing gravity at the tabletop experiments level as well as at the astrophysical observational level by deriving the corrections brought by Yukawa-like and power-law deviations from the inverse-square law. The full relativistic regime is also examined in detail.}


\begin{document}
\section{Introduction}\label{sec:I}
Bringing to light the quantum properties of gravity has recently been the subject of intense investigation. Many of the investigations conducted in the past were focused on the more modest attempt to observe the effect of gravity on a quantum particle by studying, more noticeably, the~behavior of cold neutrons inside a gravitational \mbox{field~\cite{UCNinEarth1,UCNinEarth2,UCNinEarth3,BookNeutronsInterfro,Kulin,Abele}.} In~such experiments the gravitational effect manifests itself on the quantum particle through the discrete gravitational energy spectrum the particle acquires while moving inside the gravitational field of~Earth. 

On the other hand, more recent proposals~\cite{Marletto1,Marletto2,Bose} have rather been more ambitious in the sense that they consist of experimental setups designed to make manifest the quantum character of gravity itself rather than the way a particle behaves under the influence of gravity. The~quantum character investigated is the quantum superposition of the gravitational field. Notwithstanding the higher importance of these more recent proposals towards unraveling the quantum nature of gravity, we believe that any additional effect based on the quantum interaction between the classical gravitational field and a quantum particle would, not only take us closer towards understanding gravity at the quantum level as well, but~might even give us new ways of testing classical gravity~itself.

In more recent papers, new proposals have been put forward to bring to light this interaction of a classical gravitational field with a quantum particle. In~Ref.\,\cite{COWBall}, the~possibility of probing the effects of gravity on a quantum particle in a superposition along different paths has been investigated. In~Ref.\,\cite{Josephson} the effect of gravity on a macroscopic quantum state---the superfluid made of superconducting electrons---has been investigated through the Josephson effect. In~both works, the~key idea is based on quantum interference induced by gravity (see also Refs.\,\cite{Asada1,Asada2,Asada3}). In~Ref.\,\cite{Landry1}, the~effect on the quantum states of cold neutrons of a gravitational field created by an oscillating massive object has been studied in detail. In~Ref.\,\cite{Landry2}, the~possibility of stimulating gravitons with ultra cold neutrons has been suggested. In~both of these latter works the investigation consisted of using the quantum behavior of neutrons in a known pure gravitational field---that of  Earth.
Now it would be interesting to investigate the gravitational effect on a particle, not through what is expected from its quantum behavior inside a given potential---here gravitational---but by searching for a novel effect the gravitational field could have on an otherwise usual quantum phenomena involving the particle that does not originally include gravity. Such an investigation could indeed serve two independent purposes; one which would be purely theoretical and one which might be oriented more towards the experimental side of the field. Indeed, our first purpose here is to make a new contribution to the existing proposals for making manifest the interaction of gravity with quantum particles. Our second purpose, which will be based on the outcomes of the first, is to put forward a new way for testing gravity itself using quantum~particles.

We propose here to investigate the effect of the gravitational field on the well-known Landau quantized energy levels of a charged particle moving inside a uniform and constant magnetic field (see, for~example, Ref.\,\cite{LandauLifshitz}). More specifically, we show that the gravitational field splits the Landau levels and removes their usual degeneracy (See Ref.\,\cite{EderyAudin} for a recent investigation on the effect of a linear electric field on Landau levels. In~this regard, it should be remarked here that one could conduct a similar investigation using gravity rather than an electric field. In~addition, one could then use a large massive plane with a circular hole in order to create the needed linear gravitational field that would lead to the simple harmonic oscillations of the charge in the vertical direction to the plane. The~fundamental issue behind such an approach remains, of~course, the~radiation of the charge during its simple harmonic motion in the vertical direction). It is known that this simple quantum effect helped explain the very important quantum Hall effect behind which many important technological applications have sprung (see, for~example, Refs.\,\cite{BookQHE1,BookQHE2}). 

It should be noted here, that, unlike the indifference of the quantum Hall effect to a difference in location within a gravitational field, as~shown in Ref.\,\cite{Hehl}, gravity does have an effect on the Landau levels of a charged particle as we are going to show. Our investigation aligns thus with what was pointed out in Ref.\,\cite{Grosse},  based on general grounds,  that Landau levels are split under the effect of an external potential which is monotonic in the radial distance along the plane perpendicular to the magnetic field. This does not, however, contradict what was found in Ref.\,\cite{Hehl}, as~the investigation in the latter reference was about Hall conductivity, which is indeed topological, and~hence metric-independent. As~shown in Ref.\,\cite{Hall}, the~gravitational influence on the quantum Hall effect is due to the splitting of Landau levels caused by the potential gradient of gravity in the case of Earth's field and to the quadratic gravitational potential in the case of two weakly separated massive~hemispheres.

The rest of this paper is organized as follows. In~Section~\ref{sec:II}, we derive the differential equation governing the motion of a charged particle inside a uniform magnetic field from the Klein--Gordon equation of a charged particle minimally coupled to the electromagnetic field in the curved Schwarzschild spacetime around a spherical mass. We first show how one recovers the familiar Landau levels by setting the mass to zero in the differential equation, and~then we extract an approximate equation for the case of a weak non-zero Newtonian gravitational field. In~Section~\ref{sec:III}, we expose four different methods for using the equation to extract the quantization condition. We show that only the first two methods yield a consistent and general result. The~third approach introduces an extra condition on the source of the gravitational field, whereas the last one is consistent but cannot be useful in practice. In~Section~\ref{sec:IV}, we rely on the first approach to show how to use the splitting of Landau levels induced by the gravitational field to test any departure from the square-law for gravity. We examine the widely investigated Yukawa-like and power-law potentials as concrete examples. In~Section~\ref{sec:V}, we examine the full relativistic regime of a charged particle moving within the curved Schwarzschild background under the influence of the uniform magnetic field. We conclude this paper with a brief summary and conclusion section. Appendix \ref{A} is devoted to a detailed derivation of the various mathematical formulas needed in the text.
\section{A Particle inside a Magnetic Field in the Schwarzschild~Spacetime}\label{sec:II}

Landau quantization of the energy levels of a charged particle in Minkowski spacetime arises in the presence of a uniform and constant magnetic field $\mathbf{B}$ which is perpendicular to the plane in which the charged particle is moving. In~the usual textbook treatment of the problem (see Ref.\,\cite{LandauLifshitz}) one writes down the Schr\"odinger equation for a particle, minimally interacting with the field, by~substituting the partial spatial derivatives $-i\hbar\partial_i$ by the operator $-i\hbar\partial_i-qA_i$, where $q$ is the charge of the particle and $A_i$ is the vector potential causing the magnetic field $\bf B$. 

In order to take into account the interaction of the particle with the gravitational field, however, we are going to use, instead of the Schr\"odinger equation, the~Klein--Gordon equation, $\left(\Box+m^2c^2\right)\varphi=0$. We ignore in this paper the spin of the test particle. The~equation is written in curved spacetime, where $m$ is the mass of the particle, $c$ is the speed of light, and~$\Box=g^{\mu\nu}\partial_\mu\partial_\nu$ is the d'Alembertian operator. Therefore, our particle is going to be governed by the following second-order differential equation (see Ref.\,\cite{BookQFTCS}),
\begin{equation}\label{KGCST}
\left[\frac{1}{\sqrt{-g}}D_\mu\left(\sqrt{-g}g^{\mu\nu}D_\nu\right)+m^2c^2\right]\varphi=0.    
\end{equation}

Here, $g^{\mu\nu}$ is the inverse of the spacetime metric and $g$ its determinant. We have used here the minimal prescription $D_\mu=-i\hbar\partial_\mu-eA_\mu$ to couple the particle to the four dimensional vector potential $A_\mu$ of the electromagnetic field. In~addition, in~view of the possibility of using heavy ions as test particles to minimize the contribution of the intrinsic spin, we choose here to take the electron charge $e$ as the charge of our~particle. 

We are interested in this paper in finding how the gravitational field of a spherical mass affects the Landau levels of a charged particle moving along the {equatorial plane} of the mass. For~that purpose, one needs first to consider the curved spacetime created by the spherical mass $M$. Such a spacetime is represented by the Schwarzschild metric. As~is well know in general relativity, however, one should {a priori} take into account the effect on the curvature of spacetime due to the uniform and constant magnetic field $\bf B$ as well. The~combined effect of the mass $M$ and such a $\bf B$-field---chosen to lie along the $z$ direction---gives rise to the so-called Schwarzschild--Melvin spacetime~\cite{BookExactSolutions}. It is a special case given by Ernst in Ref.\,\cite{Ernst1} of the full solution of the Einstein--Maxwell equations obtained in Ref.\,\cite{Ernst2}. In~the coordinates $(t,r,\theta,\phi)$, such a metric reads~\cite{BookExactSolutions},
\begin{equation}\label{SchwarzschildMelvin}
{\rm d}s^2=\,\Lambda^2\left(-\frac{\Delta}{r^2}\,{\rm d}t^2+\frac{r^2}{\Delta}\,{\rm d}r^2+r^2{\rm d}\theta^2\right)+\frac{r^2}{\Lambda^2}\sin^2\theta{\rm d}\phi^2.
\end{equation}

We have used here, after~restoring the standard units, the~convenient notation of Ref.\,\cite{Galtsov} in which, $\Delta=r^2-2GMr/c^2$ and $\Lambda=1+2\pi G\epsilon_0c^{-2}B^2r^2\sin^2\theta$, with~$\epsilon_0$ being the vacuum permittivity constant. On~the other hand, the~vector potential corresponding to such a constant magnetic field within the metric (\ref{SchwarzschildMelvin}) takes the form~\cite{Galtsov} (Note that we resorted back here to the covariant form of the potential vector as the tetrad form $A_{\hat{\phi}}=\frac{1}{2\Lambda}Br\sin\theta$ we displayed in our previous version of the manuscript leads to much confusion. In~fact, the~advantage of the tetrad expression displayed in the previous version of the manuscript is that it has the right dimensions for a potential vector and it allows one to straightforwardly extract the magnetic field from the spatial components using the usual curl formula, $\vec{\bf B}=\nabla\times\vec{\bf A}$),
\begin{align}\label{VectorPotential}
A_\mu=\left(0,0,0,\frac{B r^2\sin^2\theta}{2\Lambda}\right).
\end{align}

Now, the~classical motion of both neutral and charged test particles are affected by such a spacetime and have been studied analytically in detail in Ref.\,\cite{Galtsov}. It was found that the magnetic field influences neutral particles { geometrically} and creates a potential barrier that prevents both neutral and charged particles from escaping away from the axis of symmetry to infinity. In~addition, a~numerical study of the trajectories of such classical particles has also been carried out in Ref.\,\cite{Karas} with the conclusion that the particles' trajectories in such a spacetime are integrable in very special cases~only. 

In fact, unlike the Schwarzschild spacetime, the~metric (\ref{SchwarzschildMelvin}) does not reduce to the flat Minkowski metric even in the absence of the central mass $M$. In~such a case, the \mbox{metric (\ref{SchwarzschildMelvin})} represents the so-called Melvin's magnetic universe~\cite{Bonnor,Melvin}. The~investigation of the behavior of a quantum particle in Melvin's universe has been carried out in Ref.\,\cite{DiracInMelvin}. The~energy spectrum of the test particle was found to display indeed a Landau-like form which is corrected by a term arising from the {geometric} influence of the $\bf B$-field on the particle. In~addition, as~argued in Ref.\,\cite{DiracInMelvin}, such a correction term becomes significant only for a magnetic field intensity as high as $10^{19}{\rm G}$ and for a quantum number $n$ of the \mbox{order $10^{30}$.} 

However, our focus in this paper is, rather, the~effect of the gravitational field created by a spherical mass on the usual Landau levels, not the geometric, {i.e.},  the~``gravitational'', effect of the magnetic field itself.  For~that purpose, our setup here will be that of a spherical mass immersed inside a magnetic field the intensity of which brings a negligible geometric contribution to the background spacetime. Such an assumption is indeed amply sufficient for tabletop experiments on Earth as well as at the level of astrophysical objects for which the approximation $2\pi G\epsilon_0c^{-2}B^2r^2\ll1$ holds. For~young neutron stars and magnetars the intensity of the magnetic field can indeed be low and range around $10^{8\sim12}{\rm G}$ (see Ref.\,\cite{BookNeutronStar}). Under~those conditions we may thus safely adopt the metric (\ref{SchwarzschildMelvin}) with the approximation $\Lambda\sim1$, in~which case the metric reduces to the pure Schwazschild~spacetime.

Therefore, in~this paper we take the metric in Equation\,(\ref{KGCST}) to be that of the pure Schwarz-schild spacetime around a spherical mass $M$ \cite{Wald}:
\begin{equation}\label{Schwarzschild}
{\rm d}s^2=\,-\left(1-\frac{2GM}{c^2r}\right){\rm d}t^2+\left(1-\frac{2GM}{c^2r}\right)^{-1}{\rm d}r^2+r^2{\rm d}\Omega^2.
\end{equation}

Here, ${\rm d}\Omega^2={\rm d}\theta^2+\sin^2\theta{\rm d}\phi^2$ is the 2-metric on the unit sphere. Furthermore, we are going to use the vector potential $A_\mu$ of expression (\ref{VectorPotential}) which reduces, for~$\Lambda\sim1$, to~its usual form in the axially symmetric (cylindrical) gauge (the equivalent of the Landau gauge) in spherical coordinates $(t,r,\theta,\phi)$, in~which it reads, $A_\mu=(0,0,0,\tfrac{1}{2}B r^2\sin^2\theta)$. Thus, we assume, for~definiteness, that the magnetic field $\bf B$ is indeed directed along the $z$-axis.

Now, because~of the symmetry of the planar motion of the particle around the $z$-axis, we expect the wavefunction of the particle to be of the form, $\varphi(t,r,\theta,\phi)=e^{-i\frac{Et}{\hbar}}e^{i\ell\phi}R(r,\sin\theta)$, where $E$ is the energy of the particle and $\ell$ is a positive integer---we assume, for~definiteness, a~specific projection of the orbital angular momentum of the particle along the $z$-axis, {i.e.}, 
a~positive projection. Therefore, the~Klein--Gordon equation for the particle in this curved spacetime reads, explicitly,
\vspace{6pt}
\begin{multline}\label{KGinSchwR1}
\Bigg[\frac{E^2}{\hbar^2c^2}\left(1-\frac{2GM}{c^2r}\right)^{-1}-\frac{m^2c^2}{\hbar^2}+\left(\frac{2}{r}-\frac{2GM}{c^2r^2}\right)\partial_r+\left(1-\frac{2GM}{c^2r}\right)\partial_r^2\\
+\frac{\partial^2_\theta}{r^2}
+\frac{\cos\theta\partial_\theta}{r^2\sin\theta}
-\frac{\ell^2}{r^2\sin^2\theta}+\frac{eB\ell}{\hbar}
-\frac{e^2B^2r^2\sin^2\theta}{4\hbar^2}\Bigg]R(r,\theta)=0.
\end{multline}  

Furthermore, by~having the particle move along the equatorial plane, along which $\theta=\tfrac{\pi}{2}$, the~cylindrical symmetry of the system allows us to also expect the function $R(r,\theta)$ to depend only on the distance $\rho=r\sin\theta$ of the particle from the $z$-axis which is perpendicular to the plane of motion (Note that we assume here the motion of the particle to be restricted to the equatorial plane for the sake of simplicity. However, any plane perpendicular to the magnetic field would be valid provided the latitude angle $\theta\neq\frac{\pi}{2}$ of such a plane is taken into account. Around the poles of a magnetized astrophysical object, for~example, one should consider $\theta\ll1$. Indeed, we restrict here the particle's motion to a plane in view of the many applications such a configuration gives rise to. Such applications range from the two-dimensional motion of the electron gas in the quantum Hall effect~\cite{Hall} to the study of the equation of state of the dense matter confined to the thin envelop or crust of magnetized astrophysical objects, as~will be discussed in Section \ref{sec:PL}. In~fact, for~a particle which is instead free to explore the three-dimensional space the crucial Landau levels do not arise, for~then the particle's wavefunction would be described by spherical harmonics governed by a central gravitational potential inside a strong magnetic field, much like in the case of atoms inside large magnetic fields (see the discussion below Equation\,(\ref{SchwarSchrod}))). This would then make $R$ a function of the form $R(r,\sin\theta)=R(r\sin\theta)=R(\rho)$. Therefore, we have, $\partial_\theta R=r\cos\theta\partial_\rho R$. Thus, we also have, $\partial_\theta^2 R=-r\sin\theta\partial_\rho R+r^2\cos^2\theta\partial_\rho^2R$. Substituting these inside the previous equation, the~latter takes the following simplified explicit form for $\theta=\tfrac{\pi}{2}$,\\
\begin{multline}\label{KGinSchwR2}
\frac{{\rm d^2} R}{{\rm d}\rho^2}+\frac{1}{\rho}\frac{{\rm d}R}{{\rm d}\rho}+\Biggl[\frac{E^2}{\hbar^2c^2}\left(1-\frac{2GM}{c^2\rho}\right)^{-2}\\
+\left(1-\frac{2GM}{c^2\rho}\right)^{-1}\left(-\frac{m^2c^2}{\hbar^2}
-\frac{\ell^2}{\rho^2}+\frac{eB\ell}{\hbar}-\frac{e^2B^2\rho^2}{4\hbar^2}\right)\Biggr]R=0.
\end{multline}

This is the general equation that describes the planar motion of a charged particle inside a magnetic field within the spherically symmetric Schwarzschild spacetime. Before~we examine how to extract the quantization condition from this equation for weak gravitational fields, we shall first set $M=0$ and solve the equation for $R(\rho)$ in order to make contact with what we already know about the dynamics of a charged particle in a uniform magnetic field inside the Minkowski spacetime. In~fact, in~addition to allowing us to check the correctness of Equation\,(\ref{KGinSchwR2}), this first step will provide us with the fundamental wavefunction to be used later in the more interesting case of $M\neq0$.

\subsection{In Minkowski Spacetime: $M=0$}\label{sec:ForM=0}
For $M=0$, we know that we should recover the free particle of relativistic energy $E=\mathcal{E}+mc^2$, moving inside a uniform magnetic field of magnitude $B$ and perpendicular to the plane $\theta=\tfrac{\pi}{2}$. Let us assume, for~simplicity, a~non-relativistic regime for the particle, {i.e.}, $\mathcal{E}\ll mc^2$. Then, Equation\,(\ref{KGinSchwR2}) becomes, after~setting $M=0$, as~follows,
\begin{equation}\label{KGinSchwM=0}
\frac{{\rm d}^2R}{{\rm d}\rho^2}
+\frac{1}{\rho}\frac{{\rm d}R}{{\rm d}\rho}+\left(\frac{2m\mathcal{E}}{\hbar^2}+\frac{eB\ell}{\hbar}-\frac{e^2B^2\rho^2}{4\hbar^2}-\frac{\ell^2}{\rho^2}\right)R=0.
\end{equation}

In order to easily solve this equation, let us choose the following ansatz for the radial wavefunction (In the literature, the~ansatz $R(\rho)=\rho^{|\ell|}e^{-\tfrac{eB}{4\hbar}\rho^2}v(\rho)$ is sometime chosen for the radial function $R(\rho)$. This just makes  the two possibilities of a left(right)-moving particle around the mass $M$), $R(\rho)=\rho^{\ell}v(\rho)\times\exp(-\tfrac{eB}{4\hbar}\rho^2)$, and~denote by a prime a derivative with respect to $\rho$. The~equation then becomes,
\begin{equation}\label{KGinMink1}
\rho v''(\rho)+\left(2\ell+1-\beta\rho^2\right)v'(\rho)+(\alpha-\beta)\rho\, v(\rho)=0.
\end{equation}

Here, we have set, for~convenience, $\alpha=\frac{2m\mathcal{E}}{\hbar^2}$ and $\beta=\frac{eB}{\hbar}$. Next, we perform the following change of variable: $z=\tfrac{1}{2}\beta\rho^2$. This allows us, in~turn, to~rewrite the equation in the following canonical form~\cite{BookKummer},
\begin{equation}\label{KGinMink2}
zv''(z)+\left(\ell+1-z\right)v'(z)-\left(\frac{1}{2}-\frac{\alpha}{2\beta}\right)v(z)=0.
\end{equation}

This equation is of the well-known form $zv''+(b-z)v'-av=0$, called a confluent hypergeometric differential equation, the~solution of which is a linear combination of two confluent hypergeometric functions $_1F_1(a;b;z)$. These functions are also known as Kummer's functions~\cite{BookKummer}. The~general solution to the canonical Equation~(\ref{KGinMink2}) is therefore, $v(z)=A\;_1F_1(a;b;z)+A'\;z^{1-b}\,_1F_1(a-b+1;2-b;z)$. The~constants $A$ and $A'$ are the two constants of integration and, in~our case, $a=(1-\alpha/\beta)/2$ and $b=\ell+1$. Combining this general solution with our ansatz for $R(\rho)$, and~using our definition of $z$, we finally get the particle's radial wavefunction as follows,
\begin{equation}\label{WaveinMink}
R(\rho)=A\,\rho^{\ell} e^{-\frac{\beta}{4}\rho^2}\,_1F_1\left(\frac{1}{2}-\frac{\alpha}{2\beta};\ell+1;\frac{\beta}{2}\rho^2\right).
\end{equation}

We have discarded here the second solution that goes with the constant $A'$ as it would make $R(\rho)$, and~hence the wavefunction $\varphi(r)$, diverge at the origin $r=0$ for any azimuthal quantum number $\ell\geq0$. Now, keeping only this first solution (\ref{WaveinMink}), the~latter also diverges exponentially for $\rho\rightarrow\infty$ unless we impose the following condition~\cite{BookKummer,BookConfluent},
\begin{equation}\label{QuantizationinMink1}
\frac{1}{2}-\frac{\alpha}{2\beta}=-n,    
\end{equation}
for some positive integer $n$, in~which case the confluent hypergeometric function in expression (\ref{WaveinMink}) becomes a finite-degree polynomial. The~condition (\ref{QuantizationinMink1}), in~turn, implies~that,\begin{equation}\label{QuantizationinMink2}
\alpha=\beta\left(2n+1\right). 
\end{equation}

By substituting the values of $\alpha$ and $\beta$ we defined below Equation\,(\ref{KGinMink1}), we find the following more familiar quantization condition for the energy $\mathcal{E}$ of a charged particle inside a uniform magnetic field:
\begin{equation}\label{Landau}
\mathcal{E}_n=\frac{\hbar eB}{m}\left(n+\frac{1}{2}\right). 
\end{equation}

These are the usual Landau quantized energy levels in which we clearly see the high degeneracy of the levels due to the freedom the particle has with the orbital quantum number $\ell$. 
Let us now use these results to examine the case of the spherically symmetric curved spacetime in the Newtonian approximation. 
\subsection{Back to the Schwarzschild Spacetime: $M\neq0$}\label{sec:ForMNeq0}
We are interested in finding the effect of the gravitational field on the Landau \mbox{levels (\ref{Landau}).} Therefore, the~case of a spherical mass $M$ for which $GM\ll c^2\rho$ will amply be sufficient for us here. In~addition of simplifying greatly our calculations, this restriction is also greatly motivated by its practical side with regard to an eventual experimental setup. Indeed, since even for neutron stars, magnetars and magnetic white dwarfs, for~which the mass could be of the order of a few solar masses (or a fraction thereof for white dwarfs) and for which the radius ranges from a few kilometers for neutron stars/magnetars to a few thousands of kilometers for magnetic white dwarfs, the~approximation $GM\ll c^2\rho$ is very realistic and, hence, serves well our main purpose in the present~paper. 

Therefore, we can now expand what is inside the square brackets in Equation\,(\ref{KGinSchwR2}) in powers of $2GM/(c^2\rho)$ and keep only the leading order in such a ratio. Equation\,(\ref{KGinSchwR2}) then takes the following form:
\begin{multline}\label{KGinSchSmallM1}
\frac{{\rm d^2} R}{{\rm d}\rho^2}+\frac{1}{\rho}\frac{{\rm d}R}{{\rm d}\rho}+\Bigg[\frac{E^2}{\hbar^2c^2}\left(1+\frac{4GM}{c^2\rho}\right)-\frac{m^2c^2}{\hbar^2}\left(1+\frac{2GM}{c^2\rho}\right)\\
+\left(1+\frac{2GM}{c^2\rho}\right)\left(
-\frac{\ell^2}{\rho^2}+\frac{eB\ell}{\hbar}-\frac{e^2B^2\rho^2}{4\hbar^2}\right)\Bigg]R=0.
\end{multline}

This equation is very general and applicable even at the astrophysical level, provided the restrictions on the mass and the radius of the astrophysical objects of interest, as~mentioned above, are satisfied. On~the other hand, for~not too strong magnetic fields, like those used in tabletop experiments, we always have $\hbar eB/m\ll mc^2$. Furthermore, for~magnetic fields in the range $10^{3\sim9}\,{\rm G}$, like those on the surface of magnetic white dwarfs~\cite{WhiteDwarfs}, the~inequality $\hbar eB/m\ll mc^2$ holds even for electrons, whereas for neutron stars/magnetars, for~which the magnetic fields could reach the range $10^{8\sim12}\,{\rm G}$, such an inequality holds for protons and heavier ions.
Therefore, by~using $E=\mathcal{E}+mc^2$, the~non-relativistic regime for which $\hbar eB/m\ll mc^2$ holds makes the approximation $\mathcal{E}\ll mc^2$ amply sufficient for us here. This then implies that $E^2\approx 2mc^2\mathcal{E}+m^2c^4$ and Equation\,(\ref{KGinSchSmallM1}) becomes, after~keeping only the leading first-order correction coming from the $GM/(c^2\rho)$ factor, as~follows:
\begin{equation}\label{KGinSchSmallM2}
\frac{{\rm d^2} R}{{\rm d}\rho^2}+\frac{1}{\rho}\frac{{\rm d}R}{{\rm d}\rho}
+\left(\frac{2m\mathcal{E}}{\hbar^2}+\frac{eB\ell}{\hbar}+\frac{2m^2GM}{\hbar^2\rho}-\frac{e^2B^2\rho^2}{4\hbar^2}-\frac{\ell^2}{\rho^2}\right)R=0.
\end{equation}

It turns out, as~discussed in the Introduction, that there are at least four ways of using this equation to extract the quantization condition of the energy of the particle. In~what follows, we are going to present all four methods and show that only the first two are able to provide us with a consistent, practical, and~general quantization condition of~energy.
\section{Four Methods Leading to~Quantization}\label{sec:III}
Differential equations of the form similar to Equation\,(\ref{KGinSchSmallM2}) arise in many areas of both physics and quantum chemistry, ranging from the study of the harmonium to the confinement potentials~\cite{Chaudhuri,Taut,Truong,karwowski3,karwowski1,karwowski2,Caruso,karwowski4,Bouda}. Various exact and approximate methods are known for solving such an equation with central potentials~\cite{Dong1,Dong2,Ikhdir,Aguilera,Hall1,Hall4,Hall3,Hall2,SaadHall}. However, among~the more familiar ones, only two yield the right quantization condition on the energy of the particle that is both general, consistent, and~practical. We are going to examine first the ones that do yield a consistent quantization condition, but~which, unfortunately, are the least used ones in the recent physics literature dealing with the quantization condition of a particle obeying a similar equation. The~first one consists in using the time-independent perturbation theory, whereas the second one consists in approximating the system by a simple harmonic oscillator.
\subsection{Using Perturbation~Theory}\label{IIIA}
To make contact with the time-independent perturbations, we should first rewrite Equation\,(\ref{KGinSchSmallM2}) in a Schr\"odinger-like form. To~achieve that, let us set $R(\rho)=\rho^{-1/2}\psi(\rho)$.  Equation\,(\ref{KGinSchSmallM2}) then takes the form,
\begin{equation}\label{SchwarSchrod}
-\frac{\hbar^2}{2m}\psi''+\!\left(\frac{e^2B^2\rho^2}{8m}\!+\!\frac{\hbar^2(\ell^2\!-\!\tfrac{1}{4})}{2m\rho^2}\!-\!\frac{\hbar eB\ell}{2m}\!-\!\frac{GMm}{\rho}\right)\!\psi=\mathcal{E}\psi.
\end{equation}

This is just a Schr\"odinger equation with an effective potential $V_{\rm eff}(\rho)$ made of two parts. The~first part, consisting of the first three terms inside the parentheses, represents the potential of a particle inside a uniform magnetic field. This part of the effective potential is what would give rise to the familiar Landau levels (\ref{Landau}) when solving Equation\,(\ref{SchwarSchrod}) without the last term inside the parentheses. The~eigenfunctions $\psi(\rho)$ of the corresponding Hamiltonian would then be found using expression (\ref{WaveinMink}). The~last term inside the parentheses in Equation\,(\ref{SchwarSchrod}) represents the second part of the effective potential, and~constitutes just a small perturbing potential $V(\rho)$. In~fact, recall that in our approximation of a weak gravitational field, we assume that $\rho^{-1}GMm\ll e\hbar B/m$, so that the Newtonian potential $-GMm/\rho$ is indeed nothing but a small perturbation compared to the terms coming from the magnetic interaction. We are thus set for making use of the time-independent perturbation theory formalism (see Ref.\,\cite{BookQM}).

Now, since Equation\,(\ref{SchwarSchrod}) describes a particle inside a constant magnetic field and a central Newtonian potential, the~system we are dealing with is very reminiscent of the well-studied systems of atoms inside high magnetic fields (see  Refs.\,\cite{ReviewOnAtomsInStrongB,BookOnAtomsInStrongB} for a review). However, the~works dealing with atoms inside a magnetic field are concerned with finding the energy-eigenstates and the electromagnetic transitions of an atom inside a magnetic field. As~such, the~problem dealt with in the literature on the subject is that of solving for the fully three-dimensional motion of the electron moving within the central potential of the nucleus. A~specific approximation has to be used in that case (for strong magnetic fields it is called the ``adiabatic approximation'' \cite{BookOnAtomsInStrongB}), which consists in expanding the wavefunction in terms of the Landau states weighed by longitudinal wavefunctions along the direction parallel to the magnetic field. As~such, the~result is that the usual Landau levels simply get augmented by the Coulomb bound state energies as displayed in Equation\,(8.7) of Ref.\,\cite{BookOnAtomsInStrongB}. In~our case, however, we shall use instead the time-independent perturbation theory to deal with the effect of the gravitational central potential on the { planar} motion of a particle moving around a massive object inside a magnetic field. For~this reason our results will differ drastically from what is found in the case of an atom inside a strong magnetic field based on the adiabatic approximation. Our result will indeed display a product of the magnetic and gravitational contributions that has not been previously reported in the~literature.

If we denote the unperturbed $n^{\rm th}$ Landau energy level (\ref{Landau}) by $\mathcal{E}^{(0)}_n$ then, as~long as the spacing $\mathcal{E}_n^{(0)}-\mathcal{E}_m^{(0)}$ between these energy levels is greater than the Newtonian potential $GMm/\rho$, we are guaranteed that the perturbation expansion will be legitimate. The~problem that might arise with this method is that the computation might become inaccessible as the Landau levels are infinitely degenerate. Fortunately, however, the~fact that the Newtonian perturbation depends only on $\rho$, {i.e.}, is rotational symmetric, means that there is no coupling between two different Landau orbitals within the same Landau~level.

Since we already saw that only expression (\ref{WaveinMink}) converges for small $\rho$, we are going again to keep here only that expression and use it as the eigenfunction of the unperturbed Hamiltonian corresponding to Equation\,(\ref{SchwarSchrod}). Recalling then that $\psi(\rho)=\rho^{1/2}R(\rho)$, we have the following explicit expression for the unperturbed wavefunctions of the Schr\"odinger Equation~(\ref{SchwarSchrod}),
\begin{equation}
\label{UnperturbedeigenFunctions}
\psi^{(0)}_{n\ell}(\rho)=A_{n\ell}\;\rho^{\ell+\frac{1}{2}} e^{-\frac{\beta}{4}\rho^2}\,_1F_1\left(-n;\ell+1;\frac{\beta}{2}\rho^2\right),
\end{equation}
with $n=\frac{1}{2}(\alpha/\beta-1)$. Therefore, our first task is to find the normalization constants $A_{n\ell}$ which can be determined by imposing the completeness condition on the eigenfunctions, $\int_0^\infty \psi_{n\ell}^{(0)*}(\rho)\psi_{m\ell}^{(0)}(\rho) \,{\rm d}\rho=\delta_{nm}.$ However, since for a more realistic setting the sphere of mass $M$ has a finite nonzero radius $\rho_0$, much larger than its Schwarzschild radius, the~test particle's position would be limited to the interval $\rho\in[\rho_0,\infty)$. In~other words, because~of the finite size of the mass source we are considering here, the~spherical mass imposes a physical restriction on the test particle so that the interior region of the former (located below $\rho=\rho_0$) remains physically inaccessible to the latter. Iimportantly, also, is the fact that our gravitational field is valid only for $\rho\ge\rho_0$, { i.e.}, outside the spherical~mass. 

In the realistic case of a finite-radius spherical mass we should therefore distinguish two different regions when solving the Schr\"odinger equation. Region $I$, say, would represent the outside of the spherical mass, for~which $\rho>\rho_0$, while region $II$ would represent the inside of the spherical mass, for~which $\rho<\rho_0$. As~an ideal system, however, we assume the spherical mass to be completely reflective for the particle. This means the particle's wavefunction vanishes inside the sphere as the particle has no chance of penetrating inside the latter. Indeed, in~this case our system consists effectively of a particle moving around a semi-infinite potential well, inside of which the potential is infinite and outside of which the potential is just that given in Equation\,(\ref{SchwarSchrod}). The~wavefunction in the region $I$ being then just  expression (\ref{UnperturbedeigenFunctions}), all we need to impose therefore is the continuity of the latter across the surface $\rho=\rho_0$. This amounts to imposing the following requirement:
\begin{equation}\label{PsiContinuity}
    \psi^{(0)}_{n\ell}(\rho_0)=0.
\end{equation}

Based on expression (\ref{UnperturbedeigenFunctions}), this requirement, in~turn, amounts to imposing the following condition involving the confluent hypergeometric function:
\begin{equation}\label{FContinuity}
    \,_1F_1\left(-n;\ell+1;\frac{\beta}{2}\rho_0^2\right)=0.
\end{equation}

The physical meaning of condition (\ref{FContinuity}) can easily be understood as arising from the geometry of our system. In~fact, solving condition (\ref{FContinuity}) for the two unknown integers $n$ and $\ell$ simply returns these as a function of $\beta$, { i.e.}, the~magnetic field, and~the radius $\rho_0$ of the massive sphere. Indeed, by~substituting the definition of the confluent hypergeometric function given below Equation\,(\ref{PreM1ell}) into Equation\,(\ref{FContinuity}), the~latter yields a single algebraic equation involving both integers $n$ and $\ell$. By~having the finite-radius mass sit at the center does indeed geometrically disturb the motion of the particle to which all the possible Landau levels $n=1,2,3,\ldots$ and all the possible orbital numbers $\ell=1,2,3,\ldots$ would have otherwise been accessible. The~existence of the forbidden region $0<\rho\leq\rho_0$ implies that only certain values of $n$ and $\ell$ are possible depending on the value of the product $\tfrac{1}{2}\beta\rho_0^2$. Thus, to~make the particle's energies acquire the Landau levels, the~magnetic field itself should be adjusted with the size of the massive sphere to allow for the condition (\ref{FContinuity}) to be simultaneously satisfied. In~the case of a point-like mass, { i.e.}, for~$\rho_0=0$,  condition (\ref{PsiContinuity}) is automatically satisfied and one does not need to impose (\ref{FContinuity}) and, hence, no restriction is imposed on the quantum numbers $n$ and $\ell$ either. Being interested here simply in the effect of the gravitational field on the Landau energy levels, however, we are going to assume in the remainder of this paper that a specific combination of the magnetic field and the size of the sphere is already such that the existence of Landau quantum levels and orbitals is~guaranteed.

Going back to the normalization constants $A_{n\ell}$, the~normalization condition that we should impose here to get these constants is then $\int_{\rho_0}^{\infty} \psi_{n\ell}^{(0)*}(\rho)\psi_{n\ell}^{(0)}(\rho) \,{\rm d}\rho=1.$
For this purpose, we make use of the integrals derived in Appendix \ref{A}. We easily find that the constants 
$A_{n\ell}$ are given by $\mathcal{M}_{n\ell}^{-1/2}$, where $\mathcal{M}_{n\ell}$ is given by Equation\,(\ref{AppendixIntegralM}) after setting $n=m$ there.

Now, as~noted above, although~the Landau energy levels are infinitely degenerate, the~fact that the gravitational interaction is rotational symmetric means that the perturbing potential $V(\rho)$ does not couple between two different Landau orbitals of quantum numbers $\ell$ and $\ell'$. This implies that the matrix elements  $\braket{n,\ell|V(\rho)|n,\ell'}$ of the perturbation are diagonal. Therefore, the~degenerate time-independent perturbation theory yields \linebreak$\mathcal{E}_{n\ell}=\mathcal{E}_{n}^{(0)}+\braket{n,\ell|V(\rho)|n,\ell}$, where $\mathcal{E}_n^{(0)}$ is the unperturbed $n^{\rm th}$ Landau level (\ref{Landau}). This gives then the following more explicit first-order correction to the energy of the orbital $\ell$ belonging to the $n$th Landau level,
\begin{align}\label{ellCorrection}
\mathcal{E}_{n\ell}&=\mathcal{E}_{n}^{(0)}-GMm\int_{\rho_0}^{\infty} \rho^{-1}\psi^{(0)*}_{n\ell}(\rho)\psi^{(0)}_{n\ell}(\rho)\,{\rm d}\rho.
\end{align}

To evaluate the improper integral, we first substitute expressions (\ref{UnperturbedeigenFunctions}) for the wavefunctions and their normalization constants $A_{n\ell}$ as given by Equation\,(\ref{AppendixIntegralM}). Then, using the result (\ref{AppendixIntegralP}) given in Appendix \ref{A} we find,
\begin{equation}\label{1stOrderSplitting}
\mathcal{E}_{n\ell}=\mathcal{E}_{n}^{(0)}-GMm\mathcal{P}_{n\ell}\mathcal{M}_{n\ell}^{-1},
\end{equation}
where $\mathcal{P}_{n\ell}$ is given by Equation\,(\ref{AppendixIntegralP}) after setting $n=m$. Although~the product $\mathcal{P}_{n\ell}\mathcal{M}_{n\ell}$ has a long and cumbersome expression, it is actually easy to conclude from such a product that the first-order correction to the energy levels is proportional to the square root of the magnetic field. In~fact, in~both infinite series (\ref{AppendixIntegralM}) and (\ref{AppendixIntegralP}) defining $\mathcal{P}_{n\ell}$ and $\mathcal{M}_{n\ell}$, respectively, there appears the common constant factor $(\beta/2)^{-\ell}$. However, the~$\mathcal{M}_{n\ell}$-series has, in~addition, the~constant factor $(\beta/2)^{-\frac{1}{2}}$, whereas the $\mathcal{P}_{n\ell}$-series comes with the additional constant factor $(\beta/2)^{-\frac{1}{2}}$. This implies that the product $\mathcal{P}_{n\ell}\mathcal{M}_{n\ell}^{-1}$ gives rise to the constant factor $(\beta/2)^{\frac{1}{2}}$. In~light of this observation, a~better formula for the general first-order correction to the Landau levels is then the following:
\begin{equation}\label{Reduced1stOrderSplitting}
\mathcal{E}_{n\ell}=\mathcal{E}_{n}^{(0)}-GMm\sqrt{\frac{eB}{2\hbar}}\bar{\mathcal{P}}_{n\ell}\bar{\mathcal{M}}_{n\ell}^{-1}.
\end{equation}

We have introduced here the reduced series $\bar{\mathcal{P}}_{n\ell}$ and $\bar{\mathcal{M}}_{n\ell}$ which consist of expressions (\ref{AppendixIntegralP}) and (\ref{AppendixIntegralM}), respectively, without~their respective factors $(\beta/2)^{-\ell-\frac{1}{2}}$ and $(\beta/2)^{-\ell-1}$.

In order to see this more clearly, let us compute the explicit correction to the first Landau level by setting $n=1$ in Equation\,(\ref{1stOrderSplitting}). We find,
\begin{align}\label{n=1Splitting}
\mathcal{E}_{1\ell}&=\frac{3\hbar eB}{2m}-GMm\sqrt{\frac{eB}{2\hbar}}\bar{\mathcal{P}}_{1\ell}\bar{\mathcal{M}}_{1\ell}^{-1}.
\end{align}

The quantities $\mathcal{M}_{1\ell}$ and $\mathcal{P}_{1\ell}$ are given by Equations (\ref{M1ell}) and (\ref{P1ell}), respectively. This result shows that the splitting induced on the Landau levels by the gravitational field has actually a simple form at each level and for each orbital. We notice that the splitting is larger for stronger magnetic fields. The~additional term on the right-hand side of expression (\ref{n=1Splitting}) is still not fully transparent, however, as~it involves inside the product $\bar{\mathcal{P}}_{1\ell}\bar{\mathcal{M}}_{1\ell}^{-1}$ incomplete gamma functions of the form $\Gamma(\ell+1,\frac{\beta}{2}\rho_0^2)$ and $\Gamma(\ell+\frac{1}{2},\frac{\beta}{2}\rho_0^2)$, which, in~turn, involve infinite series made of $\ell$ and $\frac{\beta}{2}\rho_0^2$.

For small values of $\ell$, it is already obvious from the general definitions (\ref{AppendixIntegralMDef}) and \mbox{(\ref{AppendixIntegralPDef})} of the terms $\mathcal{M}_{1\ell}$ and $\mathcal{P}_{1\ell}$, respectively, that the product $\mathcal{P}_{1\ell}\mathcal{M}_{1\ell}^{-1}$ is of the order $\rho_0^{-1}$, so that the correction term in Equation\,(\ref{1stOrderSplitting}) is of the order $-GMm/\rho_0$, as~expected. However, in~order to extract a useful and a more transparent expression for the splitting of the $n=1$ Landau level, we shall use the large-$\ell$ limits (\ref{M1ellInfinite}) and (\ref{P1ellInfinite}) of the exact expressions (\ref{M1ell}) and (\ref{P1ell}), respectively. Then, the~splitting (\ref{n=1Splitting}) takes on the elegant and more transparent~form,
\begin{equation}\label{n=1SplittingEllInfinite}
\mathcal{E}_{1(\ell\gg1)}\approx\frac{3\hbar eB}{2m}-GMm\sqrt{\frac{eB}{2\hbar}}\,\left(\frac{\ell+\frac{3}{4}}{\ell+1}\right)\frac{\Gamma(\ell+\frac{1}{2})}{\Gamma(\ell+1)}\approx\frac{3\hbar eB}{2m}-GMm\sqrt{\frac{eB}{2\hbar\ell}}.
\end{equation}

Thus, for~large-$\ell$ orbitals the correction to the first Landau level is due to the generic product of the gravitational field contribution and the square root of the magnetic field. The~last step in Equation\,(\ref{n=1SplittingEllInfinite}) comes from the asymptotic expansion for large arguments $z$ of the gamma function~\cite{BookKummer},
\begin{equation}
    \Gamma(z)\sim z^{-\frac{1}{2}}e^{z(\log z-1)}.
\end{equation}

Therefore, the~correction term in Equation\,(\ref{n=1SplittingEllInfinite}) decreases like $1/\sqrt{\ell}$, and~thus becomes gradually suppressed for large $\ell$. In~contrast, from~Equations (\ref{AppendixIntegralM}) and (\ref{AppendixIntegralP}) giving $\mathcal{M}_{n\ell}$ and $\mathcal{P}_{n\ell}$, respectively, we see that the first-order correction (\ref{Reduced1stOrderSplitting}) does not get suppressed for large $n$. In~addition, we notice that for large $\ell$, the~correction becomes insensitive to the radius $\rho_0$ of the massive~sphere. 

We would like to emphasize here the fact that, as~explained in Appendix \ref{A}, the~large-$\ell$ approximation (\ref{n=1SplittingEllInfinite}) is valid for extremely large values of $\ell$, for~the limit was found by taking into account the already very large term $\frac{\beta}{2}\rho_0^2$ inside the incomplete gamma functions in Equations (\ref{M1ell}) and (\ref{P1ell}). Therefore, contrary to what it might seem at first sight, the~correction term on the right-hand side in Equation\,(\ref{n=1SplittingEllInfinite}) is really small (as is required for a perturbation) in comparison to the first term even for masses $M$ of the order of the solar mass. As~derived in detail in Appendix \ref{A}, this is in fact guaranteed provided that $\ell$ is bigger than $\frac{\beta\rho_0^2}{2}$. Therefore, the~correction term in Equation\,(\ref{n=1SplittingEllInfinite}) is indeed smaller than $GMm/\rho_0$, which, as~we already argued below Equation\,(\ref{SchwarSchrod}), is nothing but a small perturbation compared to the Landau energy represented by the first term. The~same remark is also valid for the general splitting formula (\ref{n=1Splitting}). In~the latter, the~smallness of the correction term compared to the first is less transparent but can, nevertheless, still be inferred from the less trivial expressions (\ref{AppendixIntegralM}) and (\ref{AppendixIntegralP}) of $\mathcal{M}_{n\ell}$ and $\mathcal{P}_{n\ell}$, respectively. In~fact, first of all, both series are exponentially suppressed by the term $e^{-\frac{\beta}{2}\rho_0^2}$. More important, however, is that the series (\ref{AppendixIntegralM}) contains the factor $\Gamma(\ell+1)$ in the numerator whereas the numerator of the series (\ref{AppendixIntegralP}) contains the factor $\Gamma(\ell+\frac{1}{2})$.

It is now important to remark here that when setting $B=0$, {i.e.}, in~the absence of the magnetic field, one does not find any quantization of energy coming from magnetism. Setting $B=0$ in Equation\,(\ref{Reduced1stOrderSplitting}), however, makes the energy vanish altogether. Actually, this is simply due to the fact that in this case both series $\mathcal{M}_{mn\ell}$ and $\mathcal{P}_{mn\ell}$ do not exist, for~the integrals that gave rise to these series vanish since the Kummer's functions vanish for $\beta=0$. A~proper treatment of the case $B=0$ consists indeed in solving the Schr\"odinger equation with the full central gravitational potential as the unique~potential. 

It is interesting to compute now the second-order correction to be able to fully appreciate the effect of the gravitational field. The~second-order corrections to the energy levels are of the form,
\begin{equation}\label{SecondOrderCorrection}
\mathcal{E}_{n\ell}^{(2)}=\sum\limits_{k\neq n}\frac{|\braket{k,\ell|V|n,\ell}|^2}{\mathcal{E}_{k}^{(0)}-\mathcal{E}_{n}^{(0)}}=\sum\limits_{k\neq n}\frac{m(GMm)^2}{2\hbar^2(k-n)}\bar{\mathcal{P}}_{kn\ell}^2\bar{\mathcal{M}}_{n\ell}^{-1}\bar{\mathcal{M}}_{k\ell}^{-1}.
\end{equation}

We have introduced here again the reduced series $\bar{\mathcal{P}}_{kn\ell}$ and $\bar{\mathcal{M}}_{n\ell}$. The~form of this correction is actually very familiar, for~one should indeed recover at the second order in $GM$ a form for the energy levels similar to that of a particle inside a central potential of the Coulombic $1/\rho$-form. This is in analogy with the hydrogen atom for which the electron's energy levels are $\propto m(k_ee^2)^2/\hbar^2$ (see, for~example, Ref.\,\cite{BookQM}). However, this expression cannot be used to find be energy levels in the case of a pure gravitational field, { i.e.}, by~setting $B=0$ inside this formula either. In~fact, a~proper treatment in this case would be to  set $B=0$ instead in Equation\,(\ref{SchwarSchrod}), as~the latter solves exactly just like for the hydrogen atom in terms of Laguerre polynomials~\cite{BookQM}.

Before we move on to the second approach, another short note is  in order here. It is actually possible to start instead from the unperturbed Hamiltonian of a particle inside the gravitational potential $V(\rho)$, together with the second term inside the parentheses in Equation\,(\ref{SchwarSchrod}), and~consider the rest of the terms rising from the interaction of the particle with the magnetic field as being the perturbing potential. This approach would also easily work because the Laguerre polynomials---which constitute then the eigenfunctions of the unperturbed Hamiltonian---are also easy to integrate like the hypergeometric functions. The~downside of this strategy for finding the effect of gravity on the Landau levels is that, experimentally, it does not make much sense to have a magnetic field so small that a mass of a few kilograms would overcome the force that such a magnetic field exerts on the charged particle. Earth's magnetic field would already affect the particle with a force that is greater than the gravitational force a few kilograms of iron would exert on the particle. Furthermore, although~the reverse might be true for astrophysical processes, it is rather still the weak gravitational field relative to the magnetic field that matters most, as~we shall discuss in Section~\ref{sec:IV}.
\subsection{Using a Harmonic Oscillator~Approximation}
This approach is based on finding the equilibrium distance of the particle from the spherical mass at which the potential energy of the particle is minimum~\cite{Taut}. In~fact, the~gravitational interaction of the particle with the spherical mass adds up to the interaction of the particle with the magnetic field to balance the centrifugal force due to the kinetic term and, hence, form the Landau bound states. This balance takes place at a specific radial distance $\rho_0$ from the center of the mass for each specific Landau level $n$ and orbital $\ell$. At~this specific radial distance, the~total potential of the particle can be approximated by that of a simple harmonic oscillator for which the quantized energy spectrum is well~known. 

Let us then start again from the Schr\"odinger Equation~(\ref{SchwarSchrod}) and expand the effective potential $V_{\rm eff}(\rho)$, contained inside the parentheses, in~a Taylor series around the equilibrium position $\rho_*$ given by $V'_{\rm eff}(\rho_*)=0$. At~the second order of the expansion, the~effective potential then reads,
\begin{equation}\label{VTaylorexpanded}
V_{\rm eff}(\rho)\simeq V_{0}+\frac{1}{2}m\omega^2(\rho-\rho_*)^2, 
\end{equation}
\textls[-5]{where $V_0=V_{\rm eff}(\rho_*)$ and $m\omega^2=V''_{\rm eff}(\rho_*)$. With~such an approximate potential, \mbox{Equation (\ref{SchwarSchrod})} takes the form of the usual Schr\"odinger equation of a simple harmonic oscillator for which the energy eigenvalues are given by,}
\begin{equation}\label{EnergySHO}
\mathcal{E}_n=V_0+\hbar\omega\left(n+\frac{1}{2}\right),
\end{equation}
where $n$ is again a positive integer. Our task now reduces then to solving for $\rho_*$ the condition $V'_{\rm eff}(\rho_*)=0$. That is, we need to solve the following quartic equation in $\rho_*$,
\begin{equation}\label{VRho0}
\frac{e^2B^2}{4m}\rho^4_*+GMm\rho_*-\frac{\hbar^2(\ell^2-\frac{1}{4})}{m}=0.
\end{equation}

Notice that the case $\ell=0$ does not arise here as in such a case, Equation\,(\ref{VRho0}) does not admit any real solution. Now, before~we solve this equation for the general case $M\neq0$ and $B\neq0$, it is instructive to examine first what would this approach give for the well-known cases of a particle inside a Coulombic potential, {i.e.}, when $B=0$, and~then for a particle inside a magnetic field only, {i.e.}, when $M=0$. This will help us find out to what extent we can rely on this approach when tackling the general case of non-vanishing $B$ and $M$. 

Setting $B=0$ in Equation\,(\ref{VRho0}) turns the latter into a first-degree equation which can easily be solved for $\rho_*$. Substituting then the resulting expression of $\rho_*$ inside Equation\,(\ref{EnergySHO}) gives the following quantized energy levels of the particle inside the gravitational field,
\begin{equation}\label{CoulombicEnergy}
\mathcal{E}_{n\ell}=\frac{m(GMm)^2}{2\hbar^2\left(\ell^2-\tfrac{1}{4}\right)^{3/2}}\left(2n+1-\sqrt{\ell^2-\tfrac{1}{4}}\right).
\end{equation}

We recognize in this expression once again the main terms characteristic of the quantized energy levels of a particle inside a Coulombic potential~\cite{LandauLifshitz,BookQM}. This would give rise to the familiar proportionality $\propto m(GMm)^2/(\hbar^2\ell^2)$ for large $\ell$. The~large-$\ell$ limit required for this approach to be accurate can be understood as enhancing our approximation of a very weak gravitational field, for~then the Lorentz force becomes indeed much bigger than the Newtonian attraction. This actually agrees qualitatively with what we found in Equation\,(\ref{SecondOrderCorrection}) using the perturbation theory approach at the second order, provided of course one takes the large-$\ell$ limit there~too.

Setting $M=0$ in Equation\,(\ref{VRho0}), on~the other hand, leaves the latter as a quartic equation, the~solution of which is, however, very easily found. Substituting then the resulting expression of $\rho_*$ inside Equation\,(\ref{EnergySHO}) gives the following quantized energy levels of the particle inside a uniform magnetic field in Minkowski spacetime,
\begin{equation}\label{RecoveringLandau}
\mathcal{E}_{n\ell}=\frac{\hbar eB}{m}\left(n+\frac{1}{2}+\frac{1}{2}\sqrt{\ell^2-\tfrac{1}{4}}-\frac{\ell}{2}\right).
\end{equation}

We recognize in the first two terms of this formula the Landau quantized energy levels. However, the~exact Formula (\ref{Landau}) is recovered only in the large-$\ell$ limit~again.

Let us now turn to the general case of a non-vanishing magnetic field in \mbox{Equation\,(\ref{VRho0}).} The~four independent solutions to general quartic equations are well known, see Ref.\,\cite{BookQuartic}. However, instead of writing down the exact cumbersome expression of the physical solution for which $\rho_*$ is real and positive, we are going to content ourselves here by extracting only an approximation for it. In~fact,
since we are already in a weak gravitational field approximation, $GMm^2\ll e^2B^2\rho_*^3$, solving exactly Equation\,(\ref{VRho0}) is not necessary for us here. Thus, the~approximate real and positive solution we find for $\rho_*$ is the following,
\begin{align}\label{Rho0}
\rho_*&\approx\sqrt{\frac{2\hbar}{eB}}\left(\ell^2-\frac{1}{4}\right)^{1/4}\left(1-x-\frac{x^2}{2}\right),\nonumber\\
x&=\frac{GMm^2}{\sqrt{8\hbar^3eB}\left(\ell^2-\frac{1}{4}\right)^{3/4}}.
\end{align}

Note that to get this approximate expression of $\rho_*$, we have relied on the additional assumption that the dimensionless parameter $x$ is very small. This assumption is amply satisfied only at laboratory-level experiments. Without~such an approximation, the~full expression of $\rho_*$ would become unwieldy and useless for neatly extracting the contribution of the gravitational field as we do in Equation\,(\ref{ApproxHOEnergyLevels}) below. In~fact, substituting expression \mbox{(\ref{Rho0})} into Equation\,(\ref{EnergySHO}), the~latter gives the sought-after quantized energy levels,
\begin{align}\label{ApproxHOEnergyLevels}
\mathcal{E}_{n\ell}&=\,\frac{\hbar eB}{m}\left(n+\frac{1}{2}+\frac{1}{2}\sqrt{\ell^2-\frac{1}{4}}-\frac{\ell}{2}\right)
+\frac{GMm}{(\ell^2-\tfrac{1}{4})^{3/4}}\sqrt{\frac{eB}{32\hbar}}\left(n+\frac{1}{2}-4\sqrt{\ell^2-\frac{1}{4}}\right)\nonumber\\
&\quad+\frac{11m(GMm)^2}{64\hbar^2(\ell^2-\tfrac{1}{4})^{3/2}}\left(n+\frac{1}{2}-\frac{8}{11}\sqrt{\ell^2-\frac{1}{4}}\right).
\end{align}

We clearly see form this expression that we recover again the usual Landau levels plus the first- and second-order corrections we obtained using perturbation theory. Both first- and second-order corrections agree qualitatively with expressions (\ref{Reduced1stOrderSplitting}) and (\ref{SecondOrderCorrection}), respectively. More importantly, however, is that for large $\ell$ the first-order correction does agree quantitatively as well with the expression obtained in Equation\,(\ref{n=1SplittingEllInfinite}) using perturbation theory. In~fact, in~addition to displaying the generic product of the gravitational field contribution and the square root of the magnetic field contribution, the~two corrections become actually identical for large $\ell$. The~numerical factors coincide and both imply a correction that decreases like $1/\sqrt{\ell}$. The~second correction does not depend on the magnetic field. It is entirely due to the the gravitational field encoded in the Schwarzschild metric, and~is quadratic in $GMm$ as in Equation\,(\ref{SecondOrderCorrection}).
\subsection{Using the Biconfluent Heun Equation: The Polynomial~Approach}
In contrast to the previous two methods for dealing with Equation\,(\ref{KGinSchSmallM2}), the~polynomial approach does not consist in extracting the energy levels from the corresponding Schr\"odinger Equation~(\ref{SchwarSchrod}) by approximately solving the latter. Instead, it is based on solving Equation\,(\ref{KGinSchSmallM2}) exactly and then requires that such a solution be physical by imposing a specific condition to be satisfied. The~quantization condition on the particle's energy thus merely comes from imposing such a condition on the~wavefunction. 

In order to solve Equation\,(\ref{KGinSchSmallM2}) exactly, we begin, as~we did in Section~\ref{sec:II}, by~setting, $R(\rho)=\rho^{\ell}v(\rho)\exp(-\tfrac{eB}{4\hbar}\rho^2)$, and~denote by a prime a derivative with respect to $\rho$. The~equation then takes the form,
\begin{equation}\label{KGinSchWithv}
\rho\, v''(\rho)+\left[2\ell+1-\beta\rho^2\right]v'(\rho)+\left[(\alpha-\beta)\rho+\gamma\right] v(\rho)=0,
\end{equation}
where, we have set, $\alpha=\frac{2m\mathcal{E}}{\hbar^2}$, $\beta=\frac{eB}{\hbar}$, and~$\gamma=\frac{2m^2GM}{\hbar^2}$. 

One way of dealing with this equation would be to expand the function $v(\rho)$ in an infinite series in the variable $\rho$. Thus, one would set $v(\rho)=\sum_{k=-\infty}^{k=+\infty} a_k\rho^k$ and one would then plug this series inside Equation\,(\ref{KGinSchWithv}). The~latter would be turned into the following identity involving an infinite series in $\rho$,
\begin{equation}\label{SchwSolutioninSeries}
\sum\limits_{k=-\infty}^{\infty}\left\{k(k+2\ell)a_k\rho^{k-1}+\gamma a_k\rho^k+\left[\alpha-(k+1)\beta\right]a_k\rho^{k+1}\right\}=0.
\end{equation}

Requiring that this identity holds for all $\rho$ leads to the following three-term \linebreak recursion relations:
\begin{equation}\label{Three-term Recursion}
(k+3)(2\ell+k+3)a_{k+3}+\gamma a_{k+2}+[\alpha-(k+2)\beta]a_{k+1}=0.
\end{equation}  

Unfortunately, there is no known general analytical solution and no simple convergence criterion for dealing with such a three-term recursion relation series. In~addition, the~truncation method one often uses to terminate the series when dealing with two-term series does not work here. A~closely related but more accurate approach, that is indeed based on a truncation method, will be given in what follows and in the following~subsection.

The right way to deal with Equation\,(\ref{KGinSchWithv}) is actually to solve exactly for the radial function $v(\rho)$ as follows. First, we need to introduce the new variable $z=\sqrt{\beta/2}\,\rho$. Substituting this inside Equation\,(\ref{KGinSchWithv}) makes the latter take the following canonical form, known as the biconfluent Heun differential equation~\cite{BookHeun,Hortacsu,Arriola,Ferreira},
\begin{equation}\label{KGinSchwInvz}
z v''(z)+\left(1+a-bz-2z^2\right)v'(z)+\left((c-a-2)z-\frac{1}{2}[d+(1+a)b]\right) v(z)=0,
\end{equation}
with $a=2\ell$, $b=0$, $c=2\alpha/\beta+2\ell$, and~$d=-2\gamma\sqrt{2/\beta}$. The~solution to this equation is given by a linear combination of two independent biconfluent Heun functions as follows,
\begin{equation}\label{KGinSchwHeunSolution}
v(z)=C_1\mathcal{H}(a,b,c,d;z)+C_2z^{-a}\mathcal{H}(-a,b,c,d;z).
\end{equation}

The constants $C_1$ and $C_2$ are the two constants of integration. Only the first term is convergent for $z\rightarrow0$, however. Although~in our case the distance $\rho$, and~hence $z$, does not go to zero because the mass used to create the Schwarzschild metric has a finite radius, we only keep the first term in order to guarantee the convergence of the series for arbitrary small, but~non-zero, $z$.

The special function $\mathcal{H}(a,b,c,d;z)$ is called the biconfluent Heun function and it is given explicitly by the following infinite series~\cite{BookHeun,karwowski4},
\begin{equation}\label{HeunFunction}
\mathcal{H}(a,b,c,d;z)=\sum\limits_{k=0}^{\infty}\frac{\mathcal{A}_k(a,b,c,d)}{(1+a)_k}\frac{z^k}{k!},
\end{equation}
where $(1+a)_k$ is the Polchhammer symbol (see Appendix \ref{A}). The~factors $\mathcal{A}_k$ in the series satisfy the following three-term recursion relation,
\vspace{6pt}
\begin{equation}\label{HeunRecursion}
    \mathcal{A}_{k+2}-\left\{(k+1)b+\tfrac{1}{2}\left[d+(1+a)b\right]\right\}\mathcal{A}_{k+1}
    +(k+1)(k+1+a)(c-a-2-2k)\mathcal{A}_k=0.
\end{equation}

Two possibilities~\cite{karwowski4} are now available for extracting valid quantization conditions on the energy of the particle based on the biconfluent Heun function (\ref{HeunFunction}). The~first will be exposed here and the second will be left for the next~subsection. 

The first possibility for finding a quantization of energy is to use the fact that the series (\ref{HeunFunction}) is highly divergent at infinity and, therefore, one necessarily needs to truncate the series to obtain an $n$-th order polynomial which would represent indeed a physical solution. For~this purpose, one easily deduces that, according to the recursion relation (\ref{HeunRecursion}), one has just to set~\cite{karwowski4},
\begin{equation}\label{TraditionalConstraint}
    c-a-2=2n\qquad\text{ and}\qquad\mathcal{A}_{n+1}=0,
\end{equation}
for some positive integer $n$. In~fact, such a requirement guarantees that all the subsequent factors $\mathcal{A}_{n+k}$ in the series (\ref{HeunRecursion}) vanish for any positive integer $k$. This method has indeed been adopted by many authors interested in finding quickly a quantization condition involving Landau levels of various systems under a magnetic field.
This approach has recently been used in Ref.\,\cite{Vieira} to study the energy levels of a galaxy moving in a Newtonian potential corrected by the cosmological constant. On~the other hand, in~Ref.\,\cite{Cunha} the effect on the Landau levels of a charged particle moving around a rotating cosmic string inside a magnetic field was also investigated by extracting a quantization rule for the energy of the charged particle based on this~approach. 

Unfortunately, the~downside of this approach is that it does not provide a consistent quantization rule for all values of the magnetic field---or of the mass-source of the gravitational field---in which the physical system is immersed. In~fact, by~substituting in the first condition of Equation\,(\ref{TraditionalConstraint}) the expressions of $a$ and $c$ as given in terms of $\alpha$, $\beta$, and~$\ell$, one just recovers a quantization condition similar to the usual Landau quantization; {i.e.}, $\alpha=\beta(n+1)$, from~which one deduces that,
\begin{equation}\label{WrongLandau}
\mathcal{E}_n=\frac{e\hbar}{2m} B(n+1).
\end{equation}

This expression is indeed similar, but~not exactly identical, to~condition
(\ref{Landau}). In~the latter the energy levels are proportional to half-integer multiples of the cyclotron frequency $eB/2\pi m$ whereas according to condition (\ref{WrongLandau}) the quantized energy is any integer or half-integer multiple of that~frequency.

Furthermore, the~second condition in Equation\,(\ref{TraditionalConstraint}) constitutes, as~already pointed out in Ref.\,\cite{karwowski4}, an~additional constraint involving again the energy $\mathcal{E}$ of the particle as well as the magnetic field $B$. In~fact, the~coefficient $\mathcal{A}_{k+1}$ in the recursion relation (\ref{HeunRecursion})
is given by the following matrix determinant~\cite{karwowski4},
\begin{equation}\label{HeunDeterminant}
\mathcal{A}_{k+1}(a,b,c,d)=
\begin{vmatrix}
\mathbb{D}_0 & 1 & 0 & 0 & 0 & \ldots & 0\\ 
\mathbb{A}_0 & \mathbb{D}_1 & 1 & 0 & 0 & \ldots & 0\\ 
0 & \mathbb{A}_1 & \mathbb{D}_2 & 1 & 0 & \ldots & 0\\ 
0 & 0 & \mathbb{A}_2 & \mathbb{D}_3 & 1 & \ldots & 0\\
\vdots & \vdots & \vdots & \ddots & \ddots & \ddots & \vdots\\
 &  &  &  & \mathbb{A}_{k-1} & \mathbb{D}_{k-1} & 1
\\
0 & 0 & 0 & 0 & 0 & \mathbb{A}_k& \mathbb{D}_k 
\end{vmatrix},
\end{equation}
where,
\begin{align}
\mathbb{D}_k&=(k+1)b+\tfrac{1}{2}\left[d+(1+a)b\right],\nonumber\\ \mathbb{A}_k&=(k+1)(k+1+a)(c-a-2-2k).
\end{align}

In our case, we have $b=0$ so that $\mathbb{D}=\tfrac{1}{2}d$. It is clear then from the matrix determinant \mbox{(\ref{HeunDeterminant})} that requiring the term $\mathcal{A}_{n+1}$ to vanish will produce an $(n+1)$-degree equation in the parameter $d=-2\gamma\sqrt{2/\beta}=-2\sqrt{eB\hbar/(2m^2GM)}$. Such an equation involves, in~addition, the~term $a=2\ell$ as well as the term $c=2\alpha/\beta+2\ell=4m\mathcal{E}/(eB\hbar)+2\ell$. 

However, having already obtained the quantization condition (\ref{WrongLandau}) that relates the energy $\mathcal{E}$ to the magnetic field $B$, it is clear that the extra $(n+1)$-degree equation $\mathcal{A}_{n+1}=0$ can only impose a specific constraint on the spherical mass $M$ used to create the gravitational field. This means that this approach cannot be consistently applied for an arbitrary gravitational field but only for cases in which one manages to fine-tune the mass $M$ in such a way to produce the quantization (\ref{WrongLandau}) itself. In~addition, such a quantization would not therefore introduce any novelty as all it yields are quantized energies similar to, but~not exactly the same as, the~usual Landau levels in Minkowski spacetime. The~only advantage of the method is thus to show that it might be possible to achieve a quantization of a particle's energy in a magnetic field even in the presence of a non-zero gravitational field by fine-tuning the mass-source of the gravitational field.
\subsection{Using the Biconfluent Heun Equation: The Asymptotic~Approach}
The asymptotic approach is again based on the solution (\ref{HeunFunction}) to Equation\,(\ref{KGinSchwInvz}) in terms of the biconfluent Heun
function. However, in~this approach one does not truncate the solution (\ref{HeunFunction}) by imposing the conditions (\ref{TraditionalConstraint}) in order to recover a finite-degree polynomial to guarantee convergence. Instead, in~this approach~\cite{karwowski4} one is rather concerned by the fact that for $z\rightarrow\infty$, the~asymptotic behavior of Heun's biconfluent function $\mathcal{H}(a,b,c,d;z)$ is given by~\cite{BookHeun,karwowski4,Temme},
\begin{equation}\label{SymptoticHeunBhavior}
\mathcal{H}(a,b,c,d;z)=\mathcal{N}(a,b,c,d)z^{-\frac{1}{2}(c+a+2)}e^{bz+z^2},
\end{equation}
where $\mathcal{N}(a,b,c,d)$ is a constant. This asymptotic behavior renders indeed the function $\mathcal{H}(a,b,c,d;x)$ not square integrable, and~hence unphysical as a wavefunction. Therefore, the~only way to guarantee square integrability, and~hence for the wavefunction to represent a physical system, is to impose $\mathcal{N}(a,b,c,d)=0$ \cite{karwowski4}. This condition is what constitutes a real quantization condition free of any inconsistency as it does not involve two separate conditions involving the energy of the particle and the parameters of the~system. 

The constant $\mathcal{N}(a,b,c,d)$ with $b=0$, which is the case of interest to us here, is given by the following infinite series~\cite{karwowski4},
\begin{equation}\label{TheNConstant}
\mathcal{N}(a,0,c,d)=\frac{\Gamma(a+1)2^{\lambda-1}}{\Gamma(a+1-\lambda)\Gamma(\lambda)}\sum\limits_{k=0}^{\infty}2^{k/2} \Gamma\left(\frac{\lambda+k}{2}\right)\mathcal{A}_k(a,0,c,d),
\end{equation}
where, $\lambda=\frac{a}{2}+\frac{c}{2}+1$ and the factors $\mathcal{A}_k(a,b,c,d)$ in the series are given by the determinant (\ref{HeunDeterminant}) in which the terms $\mathbb{D}_k$ and $\mathbb{A}_k$ are given this time by~\cite{karwowski4},
\begin{align}
\mathbb{D}_k&=\frac{\frac{c}{2}-\frac{a}{2}+k+1}{(\frac{a}{2}+\frac{c}{2}+k+2)(k+2)},\nonumber\\ \mathbb{A}_k&=\frac{d}{2\sqrt{2}(\frac{a}{2}+\frac{c}{2}+k+1)(k+1)}.
\end{align}

It is obvious from these expressions that, while the requirement $\mathcal{N}(a,0,c,d)=0$ would indeed give a genuine quantization condition, as~it involves a single equation in all the parameters of the system, it is not at all useful in practice as it requires one to find the zeros of the infinite series (\ref{TheNConstant}) in order to be able to extract the quantization~condition.

\section{Testing~Gravity}\label{sec:IV}
\subsection{With a Yukawa-Like~Deviation}\label{sec:Yukawa}
In the previous two sections we were concerned only with bringing to light the effect of a known spherically symmetric gravitational field on the familiar Landau energy levels of a charged particle inside a magnetic field. The~results we obtained in those sections may actually be put to use the other way around. In~other words, we may now use what we have learned on how to deal with the simultaneous presence of a gravitational field and a magnetic field to unravel, or~at least to test, a~given unknown spherically symmetric gravitational field based on this specific splitting of the Landau energy levels of charged~particles.

The field of testing gravity using elementary particles is also rich in intense recent investigations aimed at testing the inverse-square law for the gravitational attraction, either at short distances based on tabletop experiments~\cite{TestingGravity1,Biedermann,Kamiya}, or~at the observational astrophysical level~\cite{TestingGravity2}. The~most widely investigated form of departure from the inverse-square law for gravity has the following Yukawa-like gravitational potential (see, for~example, Refs.\,\cite{Adelberger1,Adelberger2}),
\begin{equation}\label{Yukawa}
V(\rho)=-\frac{GMm}{(1+\delta)\rho}\left(1+\delta\, e^{-\rho/\lambda}\right),    
\end{equation}
where $\delta$ is a dimensionless parameter that quantifies the relative strength of the additional Yukawa-like potential compared to the Newtonian potential, and~$\lambda$ represents the distance range of such an additional~potential.

Our task now is to use again the time-independent perturbation theory based on this new gravitational potential. Having found the contribution $\mathcal{E}_{n\ell}^{(1)}$ of the $1/\rho\,$-perturbing term in Equation\,(\ref{Reduced1stOrderSplitting}), all we need now is to evaluate the extra contribution of the perturbation coming from the second term inside the parentheses of expression (\ref{Yukawa}). More specifically, to~the first-order, the~correction to the $n^{\rm th}$ Landau level will read,
\begin{align}
\label{YukawaEllCorrection}
\mathcal{E}_{n\ell}&=\mathcal{E}_{n}^{(0){\rm L}}+(1+\delta)^{-1}\mathcal{E}_{n\ell}^{(1){\rm N}}-\delta(1+\delta)^{-1}GMm\int_{\rho_0}^{\infty} \rho^{-1}e^{-\rho/\lambda}\psi^{(0)*}_{n\ell}(\rho)\psi^{(0)}_{n\ell}(\rho)\,{\rm d}\rho,
\end{align}
where, $\mathcal{E}_{n}^{(0){\rm L}}$ is the usual unperturbed Landau $n^{\rm th}$ level, and~$\mathcal{E}_{n\ell}^{(1){\rm N}}$ is given by the second term on the right-hand side in Equation\,(\ref{Reduced1stOrderSplitting}), and~is due to the perturbation coming from the Newtonian potential. The~presence of the exponential in this integral makes its evaluation less straightforward than those needed in Section~\ref{sec:III}. The~detailed steps leading to the exact evaluation of the integral are exposed in Appendix \ref{A}. Substituting Equation\,(\ref{AppendixKummerRhoIntegral4}) into the third term of the right-hand side of Equation\,(\ref{YukawaEllCorrection}), leads to the following first-order correction to the $n$th Landau level due to the modified gravity potential (\ref{Yukawa}),
\begin{equation}
\label{YukawaEllCorrectionFinal}
\mathcal{E}_{n\ell}=\mathcal{E}_{n}^{(0){\rm L}}+(1+\delta)^{-1}\mathcal{E}_{n\ell}^{(1){\rm N}}+\delta(1+\delta)^{-1}\mathcal{E}_{n\ell}^{(1){\rm Y}},
\end{equation}
where the first-order Yukawa perturbation $\mathcal{E}_{n\ell}^{(1){\rm Y}}$ is given by the following expression:
\begin{equation}\label{YukawaCorrectionTermBlue}
\mathcal{E}_{n\ell}^{(1){\rm Y}}=-GMm\sqrt{\frac{eB}{2\hbar}}\;\bar{\mathcal{Y}}_{n\ell} \,\bar{\mathcal{M}}_{n\ell}^{-1}.
\end{equation}

We have introduced here again the reduced series $\bar{\mathcal{Y}}_{n\ell}$ and $\bar{\mathcal{M}}_{n\ell}$ which consist of expressions (\ref{AppendixIntegralY}) and (\ref{AppendixIntegralM}), respectively, but~without their respective factors $(\beta/2)^{-\ell-\frac{1}{2}}$ and $(\beta/2)^{-\ell-1}$. The~splitting of the first Landau level, $n=1$, is found by substituting expressions (\ref{Y1ell}) and (\ref{M1ell}) of $\mathcal{Y}_{1\ell}$ and $\mathcal{M}_{1\ell}$, respectively, into~(\ref{YukawaCorrectionTermBlue}).

While the result of  (\ref{YukawaCorrectionTermBlue}) has the advantage of being exact, the~fact that it involves, as~displayed in Equation\,(\ref{AppendixIntegralY}), an~infinite series in $1/\lambda$ ($\lambda$ being a very small parameter) makes it of limited practical use for estimating the order of magnitude of such a correction in a given physical setting. For~this reason, we shall, instead, look for an approximate estimate of the correction by replacing the exponential function $e^{-\rho/\lambda}$ in the integral (\ref{YukawaEllCorrection}) by the dominant value $e^{-\rho_0/\lambda}$, it takes within the whole interval of integration. In~fact, the~integrand in that integral, being already exponentially suppressed for large values of $\rho$ by the factor $e^{-\frac{\beta}{2}\rho^2}$ as we saw in Section \ref{IIIA} for the Newtonian correction, the~error we introduce in our estimate based on such a substitution will be at most of the order of $\lambda e^{-\rho_0/\lambda}$ times the exact correction. Therefore, by~performing the integral in \mbox{Equation\,(\ref{YukawaEllCorrection})} after replacing $e^{-\rho/\lambda}$ by $e^{-\rho_0/\lambda}$, we find the first-order correction to the $n$th Landau level to be given by Equation\,(\ref{YukawaEllCorrectionFinal}), where the Yukawa contribution to the correction is approximately given by,
\begin{equation}\label{YukawaCorrectionTermBlueApprox}
\mathcal{E}_{n\ell}^{(1){\rm Y}}\approx-GMme^{-\rho_0/\lambda}\sqrt{\frac{eB}{2\hbar}}\;\bar{\mathcal{P}}_{n\ell} \,\bar{\mathcal{M}}_{n\ell}^{-1}.
\end{equation}

We have again used here the results of Appendix \ref{A} where the integral in \mbox{Equation\,(\ref{YukawaEllCorrection})} without the term $e^{-\rho/\lambda}$ is found in terms of the quantities $\mathcal{M}_{1\ell}$ and $\mathcal{P}_{1\ell}$ as given by \mbox{Equations \,(\ref{M1ell}) and (\ref{P1ell}),} respectively. We introduced, as~done in Section \ref{IIIA}, the~barred quantities $\bar{\mathcal{M}}_{1\ell}$ and $\bar{\mathcal{P}}_{1\ell}$ to be able to extract the factor $\sqrt{eB/2\hbar}$. 

\subsection{With a Power-Law~Deviation}\label{sec:PL}

A widely investigated power-law departure from the inverse-square law for gravity has the following form (see, e.g.,~Ref.\,\cite{Yukawa2}),
\begin{equation}\label{PL}
V(\rho)=-\frac{GMm}{\rho}\left[1+\left(\frac{L}{\rho}\right)^s\right],    
\end{equation}
where the parameter $L$ has the dimensions of a length and quantifies thereby both the strength of the ``fifth'' force and the range of the latter. The~power $s$ is an integer which, in~the extra spatial dimensions
scenarios, is usually taken to be $s \leq 6$. Following the same steps as those we took to arrive at the result (\ref{ellCorrection}) for the inverse-square law case, we find the following more explicit first-order correction to the energy of the orbital $\ell$ belonging to the $n$th Landau level due to a power-law deviation,
\begin{align}\label{PLellCorrection}
\mathcal{E}_{n\ell}&=\mathcal{E}_{n}^{(0)}-GMm\int_{\rho_0}^{\infty}\frac{1}{\rho}\left[1+\left(\frac{L}{\rho}\right)^s\right]\psi^{(0)*}_{n\ell}(\rho)\psi^{(0)}_{n\ell}(\rho)\,{\rm d}\rho\nonumber\\
&=\mathcal{E}_{n}^{(0)}-GMm\left(\mathcal{P}_{n\ell}\mathcal{M}_{n\ell}^{-1}+L^s\mathcal{Q}_{n\ell}\mathcal{M}_{n\ell}^{-1}\right).
\end{align}

In the second line we have used the formulas given in Appendix \ref{A} to evaluate the integral. Just as we did in Section \ref{IIIA}, we shall examine here, for~concreteness, the~correction brought to the first Landau level $n=1$ and in the large-$\ell$ limit. 
Using the expressions \mbox{(\ref{M1ellInfinite}),} (\ref{P1ellInfinite}), and~(\ref{Q1ellInfinite}) for $\mathcal{M}_{1(\ell\gg1)}$, $\mathcal{P}_{1(\ell\gg1)}$, and~$\mathcal{Q}_{1(\ell\gg1)}$, respectively, from~Appendix \ref{A}, we find,
\clearpage
\begin{align}\label{PLn=1SplittingEllInfinite}
\mathcal{E}_{1(\ell\gg1)}&\approx\frac{3\hbar eB}{2m}-GMm\sqrt{\frac{eB}{2\hbar}}\Bigg[\left(\frac{\ell+\frac{3}{4}}{\ell+1}\right)\frac{\Gamma(\ell+\frac{1}{2})}{\Gamma(\ell+1)}\nonumber\\
&\qquad\qquad\qquad\qquad\qquad\qquad\qquad+L^s\left(\frac{eB}{2\hbar}\right)^{\frac{s}{2}}\left(\frac{\ell+\frac{3}{4}+\frac{s^2}{4}}{\ell+1}\right)\frac{\Gamma(\ell+\frac{1}{2}-\frac{s}{2})}{\Gamma(\ell+1)}\Bigg]\nonumber\\
&\approx\frac{3\hbar eB}{2m}-GMm\sqrt{\frac{eB}{2\hbar\ell}}\left[1+\left(\frac{eBL^2}{2\hbar\ell}\right)^{\frac{s}{2}}\right].
\end{align}

From this result we clearly see the additional contribution of the power-law deviation from the inverse-square law. In~contrast to the correction brought by a Yukawa-like deviation, the~correction introduced by a power-law deviation is not exponentially suppressed. Furthermore, the~correction is more significant, as~one might have expected, when the power $s$ is smaller. This last result can be effectively used as a means to measure the possible range of values for the parameter $L$ for a given integer power $s$.

Now, these two results can actually be used to probe any deviation from the square-law not only by using tabletop experiments, but~even by using astrophysical observations. In~fact, there is no better setup for a strong magnetic field combined with a relatively weak gravitational field required here---but still strong compared to Earth's standards---than within the environment of an astrophysical object, like a neutron star, a~magnetar, or~even a magnetic white dwarf. The~quantized Landau states of the electrons in the magnetic field of a white dwarf undergo a change in their equation of state which influences the pressure and energy density of the white dwarf. The~combined pressure and energy densities of matter and of the magnetic field determine the mass-radius relation of strongly magnetized white dwarfs~\cite{WhiteDwarfs2,WhiteDwarfs3,Mass-Radius,WhiteDwarfs4}. On~the other hand, it is known~\cite{Broderick} that the strong magnetic field of a neutron star softens the equation of state of matter inside the star also due to Landau quantization. In~addition, it was recently shown in Refs.\,\cite{Chamel1,Chamel2} that as a result of Landau quantization of electrons' motion under the strong magnetic field of a magnetar, the~neutron-drip transition in the crust of the star is shifted to either higher or lower densities depending on the magnetic field strength. The~quantization of the motion of the electrons also makes the star's crust almost incompressible, leading to a density that remains almost unchanged over a wide range of pressures. With~a splitting of the Landau levels due to the gravitational field of the star, however, we should observe a variation in any of these predicted patterns as a function of the star's mass. The~extent of such a variation should betray any eventual deviation from the inverse-square law thanks to the factor $e^{-\rho_0/\lambda}$ in Equation\,(\ref{YukawaCorrectionTermBlueApprox}) or the factor $(eBL^2/2\hbar\ell)^\frac{s}{2}$ in Equation\,(\ref{PLn=1SplittingEllInfinite}). 

Our main approximation, $\rho^{-1}GMm\ll e\hbar B/m$, and~the non-relativistic regime \linebreak$\mathcal{E}\ll mc^2$, on~which our investigations in this paper are based, easily find common ground within the environment of young neutron stars and/or magnetars and magnetized white dwarfs for which the intensity of the magnetic field is high enough compared to the gravitational interaction of the particle and, yet, is low enough for the spacetime effect of the magnetic field itself to be negligible. We thus expect our results to very well find a good application in those extreme environments of these stellar objects where both gravity and the magnetic field strengths are beyond what one could hope to achieve at the laboratory~level.

In fact, the~gravitational effect simply adds up to the already existing thermal broadening of the Landau levels~\cite{Yakovlev}. In~this case, however, the~coherence lengths in such environments are, of~course, drastically reduced. Nevertheless, as~discussed in the footnote below Equation\,(\ref{KGinSchwR1}), taking the horizontal plane to be around the poles of the star, where the longitudinal angle is $\theta\ll1$, still allows us to rely on our approach to obtain an estimate of the gravitational effect by avoiding the need for coherence extending all along the equator. Indeed, around the poles of the star the cylindrical coordinate $\rho$ is such that $\rho=r\sin\theta\ll r$. The~results obtained in Section \,\ref{IIIA} can then easily be adapted to such a case (as done in Ref.\,\cite{Hall} for the inside of a sphere) by expanding here, instead, the~non-relativistic Newtonian potential around the poles of the star into that of a simple harmonic oscillator. Denoting now the radius of the star by $R$, we have at the poles, $r=\sqrt{R^2+\rho^2}$, with~$\rho\ll R$. Therefore, we have, up~to a constant, $-GM/r\sim\frac{1}{2}GM\rho^2/R^3$. With~such a potential, Equation\,(\ref{SchwarSchrod}) can easily be solved to yield the following counterpart of \mbox{Equation\,(\ref{n=1SplittingEllInfinite})} for the splitting of the Landau levels (see Equation\,(12) of Ref.\,\cite{Hall}, where the role of the constant $K$ there is here played by $GM/R^3$):
\begin{equation}\label{AtPoles}
\mathcal{E}_{1\ell}\approx\frac{3\hbar eB}{2m}+\frac{(3+\ell)\hbar GMm}{eBR^3}.
\end{equation}

In contrast to the gravitational splitting given by Equation\,(\ref{n=1SplittingEllInfinite}) for the case of motion along the equator, the~splitting of the Landau levels for the case of motion around the poles is proportional to the orbital quantum number $\ell$ for $\ell\gg1$, and~it is inversely proportional to the magnitude $B$ of the magnetic field. The~reason being that the gravitational harmonic oscillator is stronger, the~larger the circles made by the particle around the poles are. On~the other hand, it is clear that the stronger the magnetic field, the~more overwhelmed the gravitational harmonic oscillator becomes. Remarkably, this simple non-relativistic approximation is valid at the poles regardless of the smallness of the coherence length of the particles around the poles. Note, in~addition, that, when it comes to astrophysical objects, the~uniform-magnetic field approximation we adopted throughout this paper is certainly much more adequate for the case of motion around the magnetic poles of a star than elsewhere on the surface of the latter. Now, for~the sake of generality and for wider applications, a~relativistic extension of the analysis we have done so far is needed and is going to be the subject of the next~section.

\section{The Full Relativistic~Regime}\label{sec:V}
By working within the full relativistic regime, the~full relativistic energy $E$ should be kept and, as~a consequence, terms involving products of $E$ with terms containing the radial distance $\rho$ in the general Equation~(\ref{KGinSchwR2}) cannot be eliminated. Therefore, the~equation cannot be brought to a Schr\"odinger-like form using the variable $\rho$. For~this reason, we are going to proceed in three steps. First, we set $R(\rho)=f(\rho)/\rho$ in Equation\,(\ref{KGinSchwR2}) so that the \mbox{latter becomes,}
\begin{multline}\label{KGinSchwR3}
    \frac{{\rm d^2} f}{{\rm d}\rho^2}-\frac{1}{\rho}\frac{{\rm d}f}{{\rm d}\rho}+\Bigg[\frac{E^2}{\hbar^2c^2}\left(1-\frac{2GM}{c^2\rho}\right)^{-2}\\
    +\left(1-\frac{2GM}{c^2\rho}\right)^{-1}\left(-\frac{m^2c^2}{\hbar^2}
-\frac{\ell^2}{\rho^2}+\frac{eB\ell}{\hbar}-\frac{e^2B^2\rho^2}{4\hbar^2}\right)+\frac{1}{\rho^2}\Bigg]f=0.
\end{multline}

Next, in~order to isolate the energy term $E^2$ and, at~the same time, keep the two derivative terms without any factors containing $GM/c^2\rho$, we use the Eddington--Finkelstein radial coordinate $\rho^*$ defined by~\cite{Wald}:
\begin{equation}\label{Eddington}
\rho^*=\rho+\frac{2GM}{c^2}\ln\left|\frac{c^2\rho}{2GM}-1\right|.    
\end{equation}

With such a change of the radial coordinate, Equation\,(\ref{KGinSchwR3}) becomes,
\begin{multline}\label{KGinSchwR4}
    \frac{{\rm d^2} f}{{\rm d}\rho^{*2}}-\frac{1}{\rho(\rho^*)}\frac{{\rm d}f}{{\rm d}\rho^*}+\Bigg[\frac{E^2}{\hbar^2c^2}\\
    +\left(1-\frac{2GM}{c^2\rho(\rho^*)}\right)\left(-\frac{m^2c^2}{\hbar^2}
-\frac{\ell^2-1}{\rho^2(\rho^*)}+\frac{eB\ell}{\hbar}-\frac{e^2B^2\rho^2(\rho^*)}{4\hbar^2}-\frac{2GM}{c^2\rho^3(\rho^*)}\right)\Bigg]f=0.
\end{multline}

Finally, in~order to cast this equation into a Schr\"odinger-like form,
we set \linebreak$f[\rho(\rho^*)]=\sqrt{\rho-2GM/c^2}\chi(\rho^*)$. The~equation takes then the following exact Schr\"odinger-like form (To the best of our knowledge, this equation has never been reported in the \mbox{literature before):}
\begin{align}\label{RelativisticKGinSch}
&-\frac{\hbar^2{\rm d}^2\chi}{2m{\rm d}\rho^{*2}}+\Bigg[\left(1-\frac{2GM}{c^2\rho(\rho^*)}\right)\Bigg(\frac{mc^2}{2}+\frac{\hbar^2(\ell^2-1)}{2m\rho^2(\rho^*)}-\frac{\hbar eB\ell}{2m}+\frac{e^2B^2\rho^2(\rho^*)}{8m}+\frac{\hbar^2 GM}{mc^2\rho^3(\rho^*)}\Bigg)\nonumber\\
&\qquad\qquad\qquad\qquad\qquad+\frac{3\hbar^2}{8m\rho^2(\rho^*)}{\left(1-\frac{4GM}{3c^2\rho(\rho^*)}\right)}\Bigg]\chi=\frac{E^2\chi}{2mc^2}.
\end{align}

This equation is indeed of the Schr\"odinger type and, hence, the~quantization of the eigenvalue $E^2/2mc^2$ and the corresponding perturbative corrections can easily be found using our results from Section~\ref{sec:III}. In~order to be able to use perturbation theory, we are still going to work within the approximation $2GM\ll c^2\rho(\rho^*)$.

Let us then first find the unperturbed relativistic energy $E$ by setting $M=0$ in Equation\,(\ref{RelativisticKGinSch}). In~this case, we have $\rho
^*=\rho$ and the equation takes exactly the same form as Equation\,(\ref{KGinSchwR2}) with the only difference that the term $\mathcal{E}$ on the right-hand side of Equation\,(\ref{KGinSchwR2}) is here replaced by $(E^2-m^2c^4)/2mc^2$. Therefore, the~unperturbed wavefunction $\chi^{(0)}_{n,\ell}(\rho)$ is given by expression (\ref{UnperturbedeigenFunctions}) and the unperturbed relativistic energies are quantized and \mbox{given by,}
\begin{equation}\label{UnperturbedRelE}
E_n^{(0)}=\sqrt{m^2c^4+2\hbar c^2eB\left(n+\frac{1}{2}\right)}.    
\end{equation}

This gives back the non-relativistic result (\ref{Landau}) within the approximation $\mathcal{E}\ll mc^2$. 

Now that we have the unperturbed relativistic energy, we are going to consider again Equation\,(\ref{RelativisticKGinSch}) in which we take all terms containing the ratio $GM/c^2\rho(\rho^*)$ as mere perturbing terms. Recalling, however, that our unperturbed wavefunction $\chi^{(0)}_{n,\ell}(\rho)$ depends on the variable $\rho^*$ through its dependence on $\rho$, we should re-express  Equation~(\ref{RelativisticKGinSch}) in terms of $\rho^*$ as well. For~that purpose, we need to express $\rho$ in terms of $\rho^*$. This could actually be done using Equation\,(\ref{Eddington}). However, as~the variable $\rho$ can only be extracted from \mbox{Equation\,(\ref{Eddington})} as an infinite series in $\rho^*$, and~given that $\rho$ appears in the denominator of many terms of Equation\,(\ref{RelativisticKGinSch}), we would have to truncate the infinite series by keeping only the zeroth order in $GM/c^2$. According to Equation\,(\ref{Eddington}) this amounts to simply approximating $\rho
^*$ by $\rho$. This approximation is in fact, according to Equation\,(\ref{Eddington}), valid as long as $GM\ll c^2\rho$. Keeping therefore the leading terms in $GM/c^2\rho$, Equation\,(\ref{RelativisticKGinSch}) takes the form,
\begin{multline}\label{RelativisticKGinSchLeading}
-\frac{\hbar^2}{2m}\frac{{\rm d}^2\chi}{{\rm d}\rho^{2}}+\left[\frac{e^2B^2\rho^{2}}{8m}+\frac{\hbar^2(\ell^2-\tfrac{1}{4})}{2m\rho^{2}}-\frac{\hbar eB\ell}{2m}-\frac{GM m}{\rho}\left(1-\frac{\hbar eB\ell}{m^2c^2}\right)-\frac{GMe^2B^2\rho}{4mc^2}\right]\chi\\
=\frac{E^2-m^2c^4}{2mc^2}\chi.
\end{multline}

The additional terms inside the square brackets will all supply corrections to the quantized relativistic energy (\ref{UnperturbedRelE}). Using these terms, the~first order correction to the \mbox{energy reads,}
\clearpage
\begin{align}\label{RelCorrection}
E_{n\ell}&=\Bigg(m^2c^4+2\hbar c^2eB\left(n+\frac{1}{2}\right)\nonumber\\
&\quad-\int_{\rho_0}^{\infty}\left[\frac{2GM m^2c^2}{\rho}\left(1-\frac{\hbar eB\ell}{m^2c^2}\right)+\frac{GMe^2B^2\rho}{2} \right]\chi^{(0)}_{n\ell}(\rho)\chi^{(0)}_{n\ell}(\rho)\,{\rm d}\rho\Bigg)^{\frac{1}{2}}\nonumber\\
&=\left[m^2c^4\!+\!2\hbar c^2eB\left(n+\frac{1}{2}\right)\!-\!{2GM m^2c^2}\left(1-\frac{\hbar eB\ell}{m^2c^2}\right)\mathcal{P}_{n\ell}\mathcal{M}_{n\ell}^{-1}\!-\!\frac{GMe^2B^2}{2}\mathcal{R}_{n\ell}\mathcal{M}_{n\ell}^{-1}\right]^{\frac{1}{2}}\!\!.
\end{align}

In the second line we have used again the formulas given in Appendix \ref{A} to evaluate the integral. Just as we did in Section \ref{IIIA} for the non-relativistic case, we shall examine here, for~concreteness, the~correction brought to the first Landau level $n=1$ and in the large-$\ell$ limit. In~fact, in~contrast to the general result (\ref{RelCorrection}), setting $n=1$ and letting $\ell\gg1$ has the advantage of showing explicitly the different $\ell$-dependence of both the Newtonian and the general relativistic contributions. 
Using the expressions (\ref{M1ellInfinite}), (\ref{P1ellInfinite}), and~(\ref{R1ellInfinite}) for $\mathcal{M}_{1(\ell\gg1)}$, $\mathcal{P}_{1(\ell\gg1)}$, and~$\mathcal{R}_{1(\ell\gg1)}$, respectively, from~Appendix \ref{A}, we finally find,
\begin{align}\label{RelCorrectionEllInf}
E_{1(\ell\gg 1)}
&\approx\Bigg[m^2c^4+3\hbar c^2eB-2GMm^2c^2\left(1-\frac{\hbar eB\ell}{m^2c^2}\right)\sqrt{\frac{eB}{2\hbar}}\,\left(\frac{\ell+\frac{3}{4}}{\ell+1}\right)\frac{\Gamma(\ell+\frac{1}{2})}{\Gamma(\ell+1)}\nonumber\\
&\qquad-\frac{GMe^2B^2}{ 2}\sqrt{\frac{2\hbar}{eB}}\,\left(\frac{\ell+\frac{7}{4}}{\ell+1}\right)\frac{\Gamma(\ell+\frac{3}{2})}{\Gamma(\ell+1)}\Bigg]^{1/2}\nonumber\\
&\approx\left[m^2c^4+3\hbar c^2eB-2GM m^2c^2\left(1-\frac{\hbar eB\ell}{m^2c^2}\right)\sqrt{\frac{eB}{2\hbar\ell}}-\frac{GMe^2B^2}{ 2}\sqrt{\frac{2\hbar\ell}{eB}}\right]^{1/2}\nonumber\\
&\approx\left[m^2c^4+3\hbar c^2eB-2GM m^2c^2\left(1-\frac{\hbar eB\ell}{2m^2c^2}\right)\sqrt{\frac{eB}{2\hbar\ell}}\right]^{1/2}.
\end{align}

This result gives the orbital splitting of the relativistic energy corresponding to the first Landau level for large orbitals $\ell$. Comparing this expression with the non-relativistic result (\ref{n=1SplittingEllInfinite}), we easily distinguish between the different contributions. In~fact, we see from the second step leading to Equation\,(\ref{RelCorrectionEllInf}) a distinct signature of both the special relativistic effects in the term $mc^2$ due to the fast motion of the particle as well as the general relativistic effects in the term $GM$ due to the curvature of the spacetime background. The~latter gives rise to the extra product $GMeB$ which is absent in the Newtonian approximation of the Schwarzschild metric. This simultaneous contribution of the magnetic field and the curvature of spacetime in a single term arises thanks to the angular motion of the particle within a curved background. Furthermore, we see from the last line that the larger is the mass of the charged particle, the~weaker is such a general relativistic correction. This is due to the fact that the mass-energy of the particle becomes then too large for the {coupled} magnetic and curved spacetime contribution to make much difference. The~other important fact is the different $\ell$-dependence between the Newtonian correction and the general relativistic correction. The~former scales like $1/\sqrt{\ell}$ whereas the latter scales like $\sqrt{\ell}$. This difference is due to the fact that the general relativistic setting becomes more important as the particle moves faster around the central mass. This is actually very reminiscent of the astrophysical observation that the general relativistic correction to the motion of planet Mercury is more important than those of the other planets that move slower as they are farther away from the~Sun.

This purely relativistic result (\ref{RelCorrectionEllInf}) can actually find a more immediate application for astrophysical situations like in the motion of electrons at the surface of magnetic white dwarfs and even neutron stars and/or magnetars. In~fact, for~a magnetic white dwarf of radius of about $0.015\,R_\odot$ and a mass of about $0.4\,M_\odot$ \cite{Mass-Radius}, the~ratio $GM/c^2\rho$ is of the order of $10^{-4}$ on the surface of the star. Our approximations done in this section are therefore suitable to fully apply our results (\ref{RelCorrection}) and (\ref{RelCorrectionEllInf}) to study the behavior of the charged electron gas making the surface of such stars and confront them with observations regarding the mass-radius predictions~\cite{Mass-Radius}. Furthermore, as~in the case of the Newtonian approximation done in Section \ref{IIIA}, any significant deviation from the predicted corrections (\ref{RelCorrection}) and (\ref{RelCorrectionEllInf}) within the range of our approximations would signal a possible deviation from general relativity. Of~course, our present study does not take into account the additional corrections that would come from the spin of the charged particles but, nevertheless, the~general relativistic corrections we found here would not deviate much by including the contributions of~spin.

\section{Summary and~Conclusions}\label{sec:VI}

We have studied the effect of the gravitational field of a spherical mass on a charged particle inside a uniform and constant magnetic field. We started from the full Schwarzschild metric and derived Equation\,(\ref{KGinSchSmallM1}) that describes the dynamics of the particle for both laboratory-based experiments and astrophysical level observations. Although~we subsequently simplified further the equation by keeping only the leading contribution from the Schwarzschild term $GM/(c^2\rho)$, going through Equation\,(\ref{KGinSchSmallM1}) was actually a crucial step for filtering out each individual contribution to the interaction potential. Furthermore, such a step allows us to clearly see how to adapt our present approach to other spacetime metrics as we show in Ref.\,\cite{EPJPII} and how to extend it to include relativistic regimes as we did in Section~\ref{sec:V}.

We found that the infinite Landau degeneracy is removed as the Landau orbitals of the same Landau level split in energy. We used two independent methods to achieve our goal for the case of the Newtonian approximation of the Schwarzschild metric and both gave similar results up to the approximating scheme adopted under each approach. We pointed out that a third method, which seems at first sight to be able to yield the right quantization condition, does not actually yield a complete and consistent quantization of energy. We saw in detail where the shortcoming of such an approach occurs. We also indicated in detail why a fourth method, which is capable, in~principle, of~leading us to the desired quantized energies, is not really practical for finding the right quantization~condition. 

In addition to bringing to light the effect of the gravitational field on the Landau levels, we showed that our results could also serve another purpose which is to test any departure from the inverse-square law for gravity using quantum mechanical particles. In~this paper, we restricted ourselves to the Schwarzschild spherically symmetric metric for which a Yukawa-like potential and a power-law potential are most easily incorporated. Moreover, such a spherically symmetric setup is easily achievable at the tabletop experiments level with, say, a~two-dimensional gas of electrons or protons moving around a massive sphere immersed inside a strong magnetic field. The~splitting of the Landau levels could be detected through the induced quantum Hall effect. In~fact, the~stationary states worked out in this paper find the most immediate laboratory application in experiments relying, for~example, on~the quantum Hall effect~\cite{Hall}. However, the~applications of the stationary states explored in this paper are not limited to the laboratory-scale experiments, for~the equation of state of matter making magnetized stars is studied using the Landau levels obtained using stationary states. Nevertheless, as~already indicated at the end of Section \,\ref{sec:PL}, it should be stressed here that the turbulent environment around neutron stars drastically reduces the possibility of ever observing charged particles occupying stable Landau levels around the stars. The~Schwarzschild vacuum geometry we used in this paper constitutes only a first idealized step towards a more rigorous study of, for~example, the~dynamics of the particles deep inside the stars, which require, instead, the~use of the interior Schwarzschild solution~\cite{InteriorSchwarzschild}. In~fact, the~other very interesting investigation of this gravitational splitting of Landau levels would be to consider the superfluid phases of baryonic matter inside neutron stars (see, Ref.\,\cite{SuperfluidNS}). Such a topic should be the subject of our future work that will, indeed, heavily rely on the mathematical results we obtained, and~methods we used, in~this paper. Nonetheless, the~idealized Schwarzschild geometry we started with in this paper clearly showed how gravity splits Landau levels, and~it helped us extract the various approximations that might be relevant to the matter on the surface of neutron~stars. 

At the laboratory level, however, the~Newtonian correction being of the order of $GMm/\rho_0$, a~spherical mass of the order of a thousand kilograms with a radius $\rho_0$ of one meter would only bring a splitting in the Landau levels of the order of $10^{-16}$\,eV for a proton. It should, nevertheless, be mentioned that such tiny corrections can easily be amplified if one uses instead of protons heavier quantum particles. In~fact, such a possibility of using larger and heavier quantum objects has already been contemplated in relation to quantum superposition of nano- and micron-sized quantum objects. Quantum objects consisting of molecules of masses up to $10^4$ amu have been successfully used in quantum interference experiments~\cite{10000} and much more massive quantum objects have also been investigated in the literature~\cite{104,Living,Nano1,Nano2,Nano3,Nano4,Nano5,Nano6,Nano7,Nano8}, reaching masses of up to $10^{13}\,$amu~\cite{Pino}. With~a charged molecule of such a mass, the~correction to the first Landau level reaches up to about $10^{-3}\,$eV, which is easily measurable provided that one reaches down to sufficiently low temperatures. In~fact, using low temperatures allows one not only to guarantee the longest coherence lengths possible, but~to also prevent the gravitational splitting of the Landau levels we derived here from being overwhelmed by the inevitable temperature broadening of the Landau levels. Thus, according to our formula (\ref{n=1SplittingEllInfinite}), the~threshold temperature allowed at the laboratory level is given by, $T_0\sim GMm\sqrt{eB/\hbar\ell K_B^2}$. For~test particle masses of the order of $10^{13}$\,amu, the~required temperature is of the order of $10^{-4}\,$K. On~the other hand, for~a Yukawa range $\lambda$ of the order of the micrometer~\cite{TestingGravity1,Adelberger1,Yukawa2}, the~splitting caused by a Yukawa-like deviation from the Newtonian potential would be drastically reduced below the $10^{-3}\,$eV by a factor of the order of $e^{-10^6}$. For~a power-law deviation, however, the~correction would reach $10^{-6}$ for an integer power $s=1$ and for a parameter $L$ of the order of a micrometer. Therefore, a~significant deviation from what our results predict would certainly signal a possible deviation of gravity from the inverse-square law and would even suggest the possible form of the~deviation.

We have also investigated in Section~\ref{sec:V} the full relativistic regime that allowed us to extract the energy corrections brought to a relativistic particle moving inside the curved Schwarzschild background. We saw that the different corrections neatly separate between the purely Newtonian one and the general relativistic one. We found that the corrections brought by the latter are more important for particles with large orbital quantum numbers $\ell$. Just as important is the fact that such corrections are not constrained anymore to slowly moving particles and, hence, can be confronted with observations at astrophysical levels. Indeed, as~briefly discussed in Section~\ref{sec:V}, our results are valid for the case of charged particles moving on the surface of typical magnetic white dwarfs. It should be noted, however, that despite the neat results we obtained throughout this paper for the level $n=1$ by relying on the large-$\ell$ limit, it will be of great interest to fully exploit our various results for general $n$ and $\ell$ by relying, instead, on~numerical~methods.

Now, in~addition to exploiting numerical methods, our next tasks---the first of which is already presented in Ref.\,\cite{EPJPII} for the non-relativistic case---are to (i) first consider the effect on Landau levels of more involved spacetime metrics, and~then (ii) take into account the relativistic regime into consideration and investigate closer the implications for the highly magnetized astrophysical objects. The~first allow us to use this splitting to probe in novel ways the modifications brought by GR to Newtonian gravity. The~second promises to make it possible to probe any deviation from the inverse-square law when no restriction on the magnetic strengths of the astrophysical objects is imposed, for~only within such general and hostile environments does the Yukawa correction have greater chances of being detected based on this splitting of the Landau levels. A~closer and more specialized investigation on how to adapt our approach to the case of particles with spin and how to combine the results of Section~\ref{sec:V} for the relativistic splitting of the Landau levels with astrophysical observations will be attempted~elsewhere.

\vspace{6pt} 




\funding{This research was funded by Natural Sciences and Engineering Research Council of Canada (NSERC) Discovery Grant, grant number RGPIN-2017-05388.}









\appendixtitles{yes} 
\appendixstart
\appendix
\section{Evaluating Integrals Involving Kummer's Functions Using Laguerre~Polynomials}\label{A}
In this appendix we show in detail how to evaluate the various improper integrals involving two Kummer' functions used in Sections~\ref{sec:III} and \ref{sec:IV}. Note that an integral involving two Kummer's functions might be found solely in terms of a ratio between products of gamma functions~\cite{BookKummer}. However, for~our purposes here a better strategy is to make a detour and use instead integrals involving Laguerre polynomials. In~fact, given that the gamma function $\Gamma(x)$ has simple poles at $x=0,-1,-2,\ldots$ and that our Kummer's functions come with the negative integers $-n$, plugging our completeness condition directly into the integral formulas given in Ref.\,\cite{BookKummer} would lead to ambiguous ratios between simple poles of the gamma~functions.  

Let us therefore start from the following relation between Kummer's hypergeometric function $\,_1F_1(-n,b+1;x)$ and the generalized Laguerre polynomial $L_n^{(b)}(x)$ (see \mbox{Refs.\,\cite{PaperLaguerre,BookLaguerre}),}
\begin{equation}\label{AppendixKummerLaguerre}
L_n^{(b)}(x)=\frac{\Gamma(n+b+1)}{n!\Gamma(b+1)}\,_1F_1(-n;b+1;x).
\end{equation}

On the other hand, an~improper integral, consisting of the Laplace-like transform involving two generalized Laguerre polynomials, reads,~\cite{PaperLaguerre,BookLaguerre},
\begin{equation}\label{AppendixLaguerreIntegral1}
\int_{0}^{\infty}x^{b}e^{-x}L_n^{(b)}(x)\,L_m^{(b)}(x){\rm d}x=\frac{\Gamma(n+b+1)}{n!}\delta_{nm}.
\end{equation}

Therefore, substituting the expression (\ref{AppendixKummerLaguerre}) of the generalized Laguerre polynomial in terms of Kummer's function,  also gives,
\begin{equation}\label{AppendixKummerIntegral1}
\int_{0}^{\infty}x^{b}e^{-x}\,_1F_1(-n,b+1;x)\,_1F_1(-m,b+1;x){\rm d}x\\
=\frac{n![\Gamma(b+1)]^2}{\Gamma(n+b+1)}\delta_{nm}.
\end{equation}

This result shows that our wavefunctions form a complete normalized set of basis for our eigenfunctions. 
On the other hand, we also have the following more general integral involving two generalized Laguerre polynomials~\cite{PaperLaguerre,BookLaguerre}:
\begin{multline}\label{AppendixLaguerreIntegral2}
\int_{0}^{\infty}x^{c-1}e^{-z\,x}L_n^{(b)}(z\,x)\,L_m^{(b')}(z\,x){\rm d}x\\
=z^{-c}\frac{\Gamma(c)\Gamma(n+b+1)\Gamma(m+b'+1-c)}{n!m!\Gamma(b+1)\Gamma(b'+1-c)}\,_3F_2(-n,c,c-b;b'+1,c-m-b';1).
\end{multline}

Here, $\,_3F_2(a,b,c;d,e;1)$ is the generalized hypergeometric function, and~it is defined by the following infinite series~\cite{BookKummer,BookHyperGeometric},
\begin{equation}\label{AppendixGeneralizedHyperGeo}
\,_3F_2(a,b,c;d,e;1)=\sum\limits_{k=0}^{\infty}\frac{(a)_k(b)_k(c)_k}{(d)_k(e)_kk!},
\end{equation}
\textls[-30]{where $(a)_k$ is the so-called Pochhammer symbol, defined by the product $(a)_k=a(a+1)\ldots(a+k-1)$, and~such that $(a)_0=1$. Therefore, combining Equations \,(\ref{AppendixKummerLaguerre}) and (\ref{AppendixLaguerreIntegral2}), we have also,}
\begin{multline}\label{AppendixLaguerreIntegral3}
\int_{0}^{\infty}x^{c-1}e^{-z\,x}\,_1F_1(-n;b+1;zx)\,_1F_1(-m;b'+1;zx)\,{\rm d}x\\
=z^{-c}\frac{\Gamma(c)\Gamma(b'+1)\Gamma(m+b'+1-c)}{\Gamma(b'+1-c)\Gamma(m+b'+1)}\,_3F_2(-n,c,c-b';b+1,c-m-b';1).
\end{multline}

Furthermore, we have the following useful addition theorem for Kummer's functions~\cite{BookKummer} that we are going to use:
\begin{equation}\label{AdditionTheorem}
    \,_1F_1(a,b;z+z_0)=\sum\limits_{k=0}^\infty\frac{(a)_k}{(b)_k}\frac{z_0^k}{k!}\,_1F_1(a+k,b+k;z).
\end{equation}

Therefore, by~starting from result (\ref{AppendixLaguerreIntegral3}) and shifting the integration boundary from $0$ to $x_0$ and then using theorem (\ref{AdditionTheorem}), we arrive, after~a little bit of extra work, at~the following important result,
\begin{align}\label{AppendixKummerIntegralOffset1}
&\int_{x_0}^{\infty}x^{c-1}e^{-zx}\,_1F_1(-n,b+1;zx)\,_1F_1(-m,b+1;zx){\rm d}x\nonumber\\
&=\int_{0}^{\infty}(x+x_0)^{c-1}e^{-z(x+x_0)}\,_1F_1\left[-n,b+1;z(x+x_0)\right]\,_1F_1\left[-m,b+1;z(x+x_0)\right]{\rm d}x\nonumber\\
&=e^{-zx_0}\sum\limits_{k=0}^{\infty}\sum\limits_{\substack{p=0\\q=0}}^\infty\frac{x_0^k\,\Gamma(c)}{k!\Gamma(c-k)}\frac{(-n)_p(-m)_q(zx_0)^{p+q}}{p!q!(b+1)_p(b+1)_q}\nonumber\\
&\qquad\qquad\times\int_{0}^{\infty}x^{c-k-1}e^{-zx}\,_1F_1\left(-n+p,b+1+p;zx\right)\,_1F_1\left(-m+q,b+1+q;zx\right){\rm d}x\nonumber\\
&=\sum\limits_{k=0}^{\infty}\sum\limits_{\substack{p=0\\q=0}}^\infty\frac{e^{-zx_0}\Gamma(c)(-n)_p(-m)_q\,x_0^{p+q+k}}{\,k!\,p!\,q!\,(b+1)_p(b+1)_q}\frac{\Gamma(b+1+q)\Gamma(m+b+1+k-c)}{\Gamma(b+q+1+k-c)\Gamma(m+b+1)}\,z^{p+q+k-c}\nonumber\\
&\qquad\qquad\times\,_3F_2(-n+p,c-k,c-k-b-q; b+p+1,c-k-m-b;1).
\end{align}

In the second step we have used the generalized binomial theorem to expand the term $(x+x_0)^{c-1}$ in powers of $x$ and $x_0$ for an arbitrary real exponent $c$, and~we then used theorem (\ref{AdditionTheorem}). In~the last step we have used again result (\ref{AppendixLaguerreIntegral3}). 
\subsection{Integrals Needed in Section \ref{IIIA}}
Now, from~this very general result (\ref{AppendixKummerIntegralOffset1}) we can extract all the useful integrals needed in this paper. In~fact, performing the change of variable, $x=\rho^2$, and~setting $b=\ell$, $z=\beta/2$, $c=\ell+1$, and~ $x_0=\rho_0^2$, the~above result yields,
\begin{equation}\label{AppendixIntegralMDef}
\int_{\rho_0}^{\infty}\rho^{2\ell+1}e^{-\frac{\beta}{2}\rho^2}\,_1F_1\left(-n,\ell+1;\frac{\beta}{2}\rho^2\right)\,_1F_1\left(-m,\ell+1;\frac{\beta}{2}\rho^2\right)\,{\rm d}\rho=\mathcal{M}_{mn\ell},
\end{equation}
where,
\begin{align}\label{AppendixIntegralM}
\mathcal{M}_{mn\ell}&=\sum\limits_{k=0}^{\ell}\sum\limits_{\substack{p=0\\q=0}}^\infty\frac{e^{-\frac{\beta}{2}\rho_0^2}\Gamma(\ell+1)(-n)_p(-m)_q\,\rho_0^{2(p+q+k)}}{2\,k!\,p!\,q!\,(\ell+1)_p(\ell+1)_q}\frac{\Gamma(\ell+1+q)\Gamma(m+k)}{\Gamma(q+k)\Gamma(m+\ell+1)}\nonumber\\
&\qquad\times\left(\frac{\beta}{2}\right)^{p+q+k-\ell-1}\,_3F_2(-n+p,\ell+1-k,1-k-q; \ell+p+1,1-k-m;1).
\end{align}

Notice that the series in $k$ in this last expression terminates at $\ell$ because in this case the exponent in $(x+x_0)^{c-1}$, to~which we apply the binomial theorem, is the integer $\ell$. The~result (\ref{AppendixIntegralM}) is what is needed to find the normalization constants $A_{n\ell}$ of the wavefunctions $\psi_{n\ell}(\rho)$ used in Sections~\ref{sec:III} and \ref{sec:IV}.

From Equation\,(\ref{AppendixIntegralM}), one can also find the needed expression $\mathcal{M}_{1\ell}$ by setting \linebreak$n=m=1$. Unfortunately, the~presence of the three sums in expression (\ref{AppendixIntegralM}) makes the task very tedious. Therefore, a~better strategy to get an explicit expression for $\mathcal{M}_{1\ell}$ is to just set $n=m=1$ directly in the integral (\ref{AppendixIntegralMDef}) giving $\mathcal{M}_{mn\ell}$ and then evaluate the integral. The~task becomes then very straightforward indeed and yields the following:
\begin{align}\label{PreM1ell}
\mathcal{M}_{1\ell}&=\int_{\rho_0}^{\infty}\rho^{2\ell+1}e^{-\frac{\beta}{2}\rho^2}\,_1F_1\left(-1,\ell+1;\frac{\beta}{2}\rho^2\right)\,_1F_1\left(-1,\ell+1;\frac{\beta}{2}\rho^2\right)\,{\rm d}\rho\nonumber\\
&=\int_{\rho_0}^{\infty}\rho^{2\ell+1}e^{-\frac{\beta}{2}\rho^2}\left[1-\frac{\beta}{2(\ell+1)}\rho^2\right]^2{\rm d}\rho.
\end{align}

In the second step, we used the definition, $\,_1F_1(a,b;z)=\sum_{k=0}^{\infty}\frac{(a)_k}{(b)_k}\frac{z^k}{k!}$, of~Kummer's function~\cite{BookKummer}. To~evaluate this integral (\ref{PreM1ell}), recall the definition of the incomplete gamma function~\cite{BookKummer},
\begin{equation}
\int_{t_0}^{\infty}t^{s-1}e^{-t}{\rm d}t=\Gamma(s,t_0),
\end{equation}
from which the variable re-definitions, $t=\frac{\beta}{2}\rho^2$ and $t_0=\frac{\beta}{2}\rho_0^2$, yield,
\begin{align}\label{IncompleteGamma}
\int_{\rho_0}^{\infty}\rho^{2s-1}e^{-\frac{\beta}{2}\rho^2}{\rm d}\rho&=\frac{2^{s-1}}{\beta^s}\Gamma\left(s,\tfrac{\beta}{2}\rho_0^2\right)\nonumber\\
&=\frac{2^{s-1}}{\beta^s}\left[\Gamma(s)-\sum\limits_{n=0}^{\infty}\frac{(-1)^n}{n!}\frac{(\frac{\beta}{2}\rho_0^2)^{s+n}}{s+n}\right].
\end{align}

In the second line we have used the explicit expression of the incomplete gamma function in terms of a sum of the usual gamma function and an infinite series~\cite{BookKummer}. Expanding the square brackets of the integral (\ref{PreM1ell}) and then applying the result (\ref{IncompleteGamma}) to each term of that sum individually, we easily find the sought-after expression,
\begin{equation}\label{M1ell}
\mathcal{M}_{1\ell}=\frac{2^{\ell}}{\beta^{\ell+1}}\Bigg[\Gamma(\ell+1,\tfrac{\beta}{2}\rho_0^2)-\frac{\Gamma(\ell+2,\tfrac{\beta}{2}\rho_0^2)}{(\ell+1)/2}+\frac{\Gamma(\ell+3,\tfrac{\beta}{2}\rho_0^2)}{(\ell+1)^2}\Bigg].
\end{equation}

As explained in the text below Equation\,(\ref{n=1Splitting}), we are interested in the large-$\ell$ limit of $\mathcal{M}_{1\ell}$. From~expression (\ref{M1ell}) it is actually easy to estimate $\mathcal{M}_{1\ell}$ in the large-$\ell$ limit using the explicit expression of the incomplete gamma function as given by the square brackets in the second line of Equation\,(\ref{IncompleteGamma}) as well as the recursion relation $\Gamma(x+1)=x\Gamma(x)$. In~fact, we have,
\begin{align}\label{M1ellInfinite}
\mathcal{M}_{1(\ell\gg1)}&\approx\frac{2^{\ell}}{\beta^{\ell+1}}\left[ \Gamma(\ell+1)-\frac{2\Gamma(\ell+2)}{\ell+1}+\frac{\Gamma(\ell+3)}{(\ell+1)^2}\right]
\approx\frac{2^{\ell}}{\beta^{\ell+1}}\frac{\Gamma(\ell+1)}{\ell+1}.
\end{align}

Note that to get to the middle step in Equation\,(\ref{M1ellInfinite}), we had to discard the term $\frac{\beta}{2}\rho_0^2$ from the incomplete gamma functions in Equation\,(\ref{M1ell}). This becomes possible only for values of $\ell$ bigger than the dimensionless factor $\frac{\beta}{2}\rho_0^2$, as~can be seen by examining the infinite defining series of the incomplete gamma function on the right-hand side of \mbox{Equation\,(\ref{IncompleteGamma}).} In~fact, for~the ratio $\frac{1}{s}\left(\beta\rho_0^2/2\right)^s$ of the infinite series inside the square brackets to be negligible compared to the first term $\Gamma(s)\sim s!$, one needs to have $s>\beta\rho_0^2e/2$, where $e$---not to be confused with the electron's charge---is the base of the natural logarithm. This condition is arrived at by evaluating the limit of the ratio $\frac{1}{s!}\left(\beta\rho_0^2/2\right)^s$ using Stirling's approximation formula for large $n$: $\ln n!\sim n\ln n-n$.    Thus, $\ell$ needs to be bigger than $\beta\rho_0^2e/2$ for the approximation (\ref{M1ellInfinite}) to hold. The~last term in expression (\ref{M1ellInfinite}) is used in Section~\ref{sec:III} to estimate the splitting of the energy levels for large orbitals $\ell$. 

Another useful integral can be extracted from the general result (\ref{AppendixKummerIntegralOffset1}) by performing again the change of variable $x=\rho^2$ and by setting this time $b=\ell$, $z=\beta/2$, $c=\ell+\frac{1}{2}$ and $x_0=\rho_0^2$. We easily find then the following result:
\begin{equation}\label{AppendixIntegralPDef}
\int_{\rho_0}^{\infty}\rho^{2\ell}e^{-\frac{\beta}{2}\rho^2}\,_1F_1\left(-n,\ell+1;\frac{\beta}{2}\rho^2\right)\,_1F_1\left(-m,\ell+1;\frac{\beta}{2}\rho^2\right)\,{\rm d}\rho=\mathcal{P}_{mn\ell},
\end{equation}
where,
\begin{align}\label{AppendixIntegralP}
\mathcal{P}_{mn\ell}&=\sum\limits_{k=0}^{\infty} \sum\limits_{\substack{p=0\\q=0}}^\infty\frac{e^{-\frac{\beta}{2}\rho_0^2}\Gamma(\ell+\frac{1}{2})(-n)_p(-m)_q\,\rho_0^{2(p+q+k)}}{2\,k!\,p!\,q!\,(\ell+1)_p(\ell+1)_q}\frac{\Gamma(\ell+q+1)\Gamma(m+\frac{1}{2}+k)}{\Gamma(q+k+\frac{1}{2})\Gamma(m+\ell+1)}\nonumber\\
&\qquad\times\left(\frac{\beta}{2}\right)^{p+q+k-\ell-\frac{1}{2}}\,\,_3F_2(-n+p,\ell+\tfrac{1}{2}-k,\tfrac{1}{2}-k-q; \ell+p+1,\tfrac{1}{2}-k-m;1).
\end{align}

This result is what allows us to find in Section~\ref{sec:III} the correction to the quantized energy levels $\mathcal{E}_n$ due to the splitting of the orbitals. We would like to find from this result the expression of $\mathcal{P}_{1\ell}$ by setting $n=m=1$. However, like for the expression $\mathcal{M}_{1\ell}$, it is also here much easier instead to proceed by substituting $n=m=1$ directly into the integral \mbox{(\ref{AppendixIntegralPDef})} defining $\mathcal{P}_{mn\ell}$ and then evaluate the former. In~fact, following similar steps as those that led us to the result (\ref{M1ell}), we easily find the following expression for $\mathcal{P}_{1\ell}$,
\begin{align}\label{P1ell}
\mathcal{P}_{1\ell}&=\int_{\rho_0}^{\infty}\rho^{2\ell}e^{-\frac{\beta}{2}\rho^2}\,_1F_1\left(-1,\ell+1;\frac{\beta}{2}\rho^2\right)\,_1F_1\left(-1,\ell+1;\frac{\beta}{2}\rho^2\right)\,{\rm d}\rho\nonumber\\
&=\int_{\rho_0}^{\infty}\rho^{2\ell}e^{-\frac{\beta}{2}\rho^2}\left[1-\frac{\beta}{2(\ell+1)}\rho^2\right]^2{\rm d}\rho\nonumber\\
&=\frac{2^{\ell-\frac{1}{2}}}{\beta^{\ell+\frac{1}{2}}}\Bigg[\Gamma(\ell+\tfrac{1}{2},\tfrac{\beta}{2}\rho_0^2)-\frac{\Gamma(\ell+\tfrac{3}{2},\tfrac{\beta}{2}\rho_0^2)}{(\ell+1)/2}+\frac{\Gamma(\ell+\tfrac{5}{2},\tfrac{\beta}{2}\rho_0^2)}{(\ell+1)^2}\Bigg].
\end{align}

From this expression it is also easy to estimate $\mathcal{P}_{1\ell}$ in the large-$\ell$ limit, as~we did for the term $\mathcal{M}_{1\ell}$ above, using the explicit expression of the incomplete gamma function as given by the square brackets in the second line of Equation\,(\ref{IncompleteGamma}) as well as the recursion relation $\Gamma(x+1)=x\Gamma(x)$. In~fact, we have,
\begin{align}\label{P1ellInfinite}
\mathcal{P}_{1(\ell\gg1)}&\approx\frac{2^{\ell-\frac{1}{2}}}{\beta^{\ell+\frac{1}{2}}}\left[ \Gamma(\ell+\tfrac{1}{2})-\frac{2\Gamma(\ell+\tfrac{3}{2})}{\ell+1}+\frac{\Gamma(\ell+\tfrac{5}{2})}{(\ell+1)^2}\right]\approx\frac{2^{\ell-\frac{1}{2}}}{\beta^{\ell+\frac{1}{2}}}\frac{(\ell+\tfrac{3}{4})\Gamma(\ell+\tfrac{1}{2})}{(\ell+1)^2}.
\end{align}

The same remark, as~the one made below Equation\,(\ref{M1ellInfinite}), about how large should $\ell$ be for the middle step in the approximation (\ref{P1ellInfinite}) to be justified, holds here also.
\subsection{Integrals Needed in Section \ref{sec:Yukawa}}
The other category of integrals we need to evaluate now is the one of Section \ref{sec:Yukawa}. The~integral reads,
\begin{align}\label{AppendixIntegralForYukawa1}
&\int_{\rho_0}^{\infty}\rho^{2\ell}e^{-\frac{\beta}{2}\rho^2-\frac{\rho}{\lambda}}\,_1F_1\left(-n,\ell+1;\frac{\beta}{2}\rho^2\right)\,_1F_1\left(-m,\ell+1;\frac{\beta}{2}\rho^2\right){\rm d}\rho\nonumber\\
&=\sum\limits_{r=0}^{\infty}\frac{(-1)^r}{r!\lambda^r}\int_{\rho_0}^{\infty}\rho^{2\ell+r}e^{-\frac{\beta}{2}\rho^2}\,_1F_1\left(-n,\ell+1;\frac{\beta}{2}\rho^2\right)\,_1F_1\left(-m,\ell+1;\frac{\beta}{2}\rho^2\right){\rm d}\rho.
\end{align}

In the second line we have the Taylor-expanded  exponential $\exp(-\rho/\lambda)$. The~integral can be evaluated using the general result (\ref{AppendixKummerIntegralOffset1}) by performing the change of variable $x=\rho^2$ and by setting this time $b=\ell$, $z=\beta/2$, $c=\ell+\frac{r}{2}+\frac{1}{2}$, and~$x_0=\rho_0^2$. We easily arrive then at the following:
\begin{equation}\label{AppendixIntegralYDef}
\int_{\rho_0}^{\infty}\rho^{2\ell}e^{-\frac{\beta}{2}\rho^2-\frac{\rho}{\lambda}}\,_1F_1\left(-n,\ell+1;\frac{\beta}{2}\rho^2\right)\\
\hfill\times\,_1F_1\left(-m,\ell+1;\frac{\beta}{2}\rho^2\right){\rm d}\rho=\mathcal{Y}_{mn\ell},
\end{equation}
where,
\begin{multline}\label{AppendixIntegralY}
\!\!\!\!\!\!\!\mathcal{Y}_{mn\ell}=\sum\limits_{\substack{k=0\\r=0}}^{\infty}\sum\limits_{\substack{p=0\\q=0}}^\infty\frac{e^{-\frac{\beta}{2}\rho_0^2}(-1)^r\Gamma(\ell+\frac{r}{2}+\frac{1}{2})(-n)_p(-m)_q\,\rho_0^{2(p+q+k)}}{2r!\,k!\lambda^r\,p!\,q!\,(\ell+1)_p(\ell+1)_q}\\\times\frac{\Gamma(\ell+1+q)\Gamma(m+\frac{1}{2}+k-\frac{r}{2})}{\Gamma(q+\frac{1}{2}+k-\frac{r}{2})\Gamma(m+\ell+1)}\,\left(\frac{\beta}{2}\right)^{p+q+k-\ell-\frac{r}{2}-\frac{1}{2}}\\
\times\,_3F_2(-n+p,\ell+\tfrac{r}{2}+\tfrac{1}{2}-k,\tfrac{r}{2}+\tfrac{1}{2}-k-q;\ell+p+1,\tfrac{r}{2}+\tfrac{1}{2}-k-m;1).
\end{multline}

To the best of our knowledge, these integrals that are shifted from the origin and that involve, in~addition, an~extra exponential function have not been previously given in the~literature. 

Now, just as we did for the previous integrals, we would also like to evaluate the quantity $\mathcal{Y}_{1\ell}$ by setting $m=n=1$. As~with the extraction of the quantities (\ref{M1ell}) and (\ref{P1ell}), it is very tedious to attempt to use the general result (\ref{AppendixIntegralY}). Therefore, we are instead going to substitute again $m=n=1$ directly into the defining integral (\ref{AppendixIntegralYDef}) of $\mathcal{Y}_{mn\ell}$. Following exactly the same steps used to get $\mathcal{M}_{1\ell}$ and $\mathcal{P}_{1\ell}$, we find,
\begin{align}\label{Y1ell}
\mathcal{Y}_{1\ell}&=\sum\limits_{r=0}^{\infty}\frac{(-1)^r}{r!\lambda^r}\int_{\rho_0}^{\infty}\rho^{2\ell+r}e^{-\frac{\beta}{2}\rho^2}\,_1F_1\left(-1,\ell+1;\frac{\beta}{2}\rho^2\right)\,_1F_1\left(-1,\ell+1;\frac{\beta}{2}\rho^2\right)\,{\rm d}\rho\nonumber\\
&=\sum\limits_{r=0}^{\infty}\frac{(-1)^r}{r!\lambda^r}\int_{\rho_0}^{\infty}\rho^{2\ell+r}e^{-\frac{\beta}{2}\rho^2}\left[1-\frac{\beta}{2(\ell+1)}\rho^2\right]^2{\rm d}\rho\nonumber\\
&=\sum\limits_{r=0}^{\infty}\frac{(-1)^r}{r!\lambda^r}\frac{2^{\ell+\tfrac{r}{2}-\frac{1}{2}}}{\beta^{\ell+\frac{r}{2}+\frac{1}{2}}}\Bigg[\Gamma(\ell+\tfrac{r}{2}+\tfrac{1}{2},\tfrac{\beta}{2}\rho_0^2)
-\frac{\Gamma(\ell+\tfrac{r}{2}+\tfrac{3}{2},\tfrac{\beta}{2}\rho_0^2)}{(\ell+1)/2}+\frac{\Gamma(\ell+\tfrac{r}{2}+\tfrac{5}{2},\tfrac{\beta}{2}\rho_0^2)}{(\ell+1)^2}\Bigg].
\end{align}
\subsection{Integral Needed in Section \ref{sec:PL}}

The second integral displayed in Section \ref{sec:PL} is of the form,
\begin{equation}\label{AppendixIntegralQDef}
\int_{\rho_0}^{\infty}\rho^{2\ell-s}e^{-\frac{\beta}{2}\rho^2}\,_1F_1\left(-n,\ell+1;\frac{\beta}{2}\rho^2\right)\,_1F_1\left(-m,\ell+1;\frac{\beta}{2}\rho^2\right)\,{\rm d}\rho\equiv\mathcal{Q}_{mn\ell}.
\end{equation}

This integral can be performed using the very general result (\ref{AppendixKummerIntegralOffset1}) by performing the change of variable, $x=\rho^2$, and~by setting $b=\ell$, $z=\beta/2$, $c=\ell+\frac{1}{2}-\tfrac{s}{2}$, and~ $x_0=\rho_0^2$. The~result is,

\begin{multline}\label{AppendixIntegralQ}
\mathcal{Q}_{mn\ell}=\sum\limits_{k=0}^{\infty} \sum\limits_{\substack{p=0\\q=0}}^\infty\frac{e^{-\frac{\beta}{2}\rho_0^2}\Gamma(\ell+\frac{1}{2}-\frac{s}{2})(-n)_p(-m)_q\,\rho_0^{2(p+q+k)}}{2\,k!\,p!\,q!\,(\ell+1)_p(\ell+1)_q}\\
\times\frac{\Gamma(\ell+q+1)\Gamma(m-\frac{s}{2}-\frac{1}{2}+k)}{\Gamma(q+k-\frac{s}{2})\Gamma(m+\ell+1)}\left(\frac{\beta}{2}\right)^{p+q+k-\ell-\frac{1}{2}+\frac{s}{2}}\\
\times\,_3F_2(-n+p,\ell+\tfrac{1}{2}-\tfrac{s}{2}-k,\tfrac{1}{2}-\tfrac{s}{2}-k-q; \ell+p+1,\tfrac{1}{2}-\tfrac{s}{2}-k-m;1).
\end{multline}

Next, by~following the same steps that led us to expression (\ref{M1ell}) for $\mathcal{M}_{1\ell}$ and to expression (\ref{P1ell}) for $\mathcal{P}_{1\ell}$, we easily find,
\begin{equation}\label{Q1ell}
\mathcal{Q}_{1\ell}=\frac{2^{\ell-\frac{1}{2}-\frac{s}{2}}}{\beta^{\ell+\frac{1}{2}-\frac{s}{2}}}\Bigg[\Gamma(\ell+\tfrac{1}{2}-\tfrac{s}{2},\tfrac{\beta}{2}\rho_0^2)-\frac{\Gamma(\ell+\frac{3}{2}-\tfrac{s}{2},\tfrac{\beta}{2}\rho_0^2)}{(\ell+1)/2}+\frac{\Gamma(\ell+\frac{5}{2}-\tfrac{s}{2},\tfrac{\beta}{2}\rho_0^2)}{(\ell+1)^2}\Bigg].
\end{equation}

Being interested in the large-$\ell$ limit of $\mathcal{Q}_{1\ell}$, we find in analogy with (\ref{M1ellInfinite}) and \mbox{(\ref{P1ellInfinite}) that,}
\begin{align}\label{Q1ellInfinite}
\mathcal{Q}_{1(\ell\gg1)}&\approx\frac{2^{\ell-\frac{1}{2}-\frac{s}{2}}}{\beta^{\ell+\frac{1}{2}-\frac{s}{2}}}\left[ \Gamma(\ell+\tfrac{1}{2}-\tfrac{s}{2})-\frac{2\Gamma(\ell+\frac{3}{2}-\tfrac{s}{2})}{\ell+1}+\frac{\Gamma(\ell+\frac{5}{2}-\tfrac{s}{2})}{(\ell+1)^2}\right]\nonumber\\
&\approx\frac{2^{\ell-\frac{1}{2}-\frac{s}{2}}}{\beta^{\ell+\frac{1}{2}-\frac{s}{2}}}\frac{(\ell+\frac{3}{4}+\frac{s^2}{4})\Gamma(\ell+\frac{1}{2}-\tfrac{s}{2})}{(\ell+1)^2}.
\end{align}

The right-hand side of the last equality is used in Section \ref{sec:PL} to estimate the splitting of the energy levels for large orbitals $\ell$.

\subsection{Integral Needed in Section~\ref{sec:V}}

The second integral displayed in Section~\ref{sec:V} is of the form,
\begin{equation}\label{AppendixIntegralRDef}
\int_{\rho_0}^{\infty}\rho^{2\ell+2}e^{-\frac{\beta}{2}\rho^2}\,_1F_1\left(-n,\ell+1;\frac{\beta}{2}\rho^2\right)\,_1F_1\left(-m,\ell+1;\frac{\beta}{2}\rho^2\right)\,{\rm d}\rho\equiv\mathcal{R}_{mn\ell}.
\end{equation}

This integral can be performed using the very general result (\ref{AppendixKummerIntegralOffset1}) by performing the change of variable, $x=\rho^2$, and~by setting $b=\ell$, $z=\beta/2$, $c=\ell+\tfrac{3}{2}$, and~ $x_0=\rho_0^2$. The~result is,
\begin{align}\label{AppendixIntegralR}
\mathcal{R}_{mn\ell}&=\sum\limits_{k=0}^{\infty} \sum\limits_{\substack{p=0\\q=0}}^\infty\frac{e^{-\frac{\beta}{2}\rho_0^2}\Gamma(\ell+\frac{3}{2})(-n)_p(-m)_q\,\rho_0^{2(p+q+k)}}{2\,k!\,p!\,q!\,(\ell+1)_p(\ell+1)_q}\frac{\Gamma(\ell+q+1)\Gamma(m-\frac{1}{2}+k)}{\Gamma(q+k-\frac{1}{2})\Gamma(m+\ell+1)}\nonumber\\
&\qquad\times\left(\frac{\beta}{2}\right)^{p+q+k-\ell-\frac{3}{2}}\,\,_3F_2(-n+p,\ell+\tfrac{3}{2}-k,\tfrac{3}{2}-k-q; \ell+p+1,\tfrac{3}{2}-k-m;1).
\end{align}

Next, by~following the same steps that led us to expression (\ref{M1ell}) for $\mathcal{M}_{1\ell}$ and to expression (\ref{P1ell}) for $\mathcal{P}_{1\ell}$, we easily find,
\begin{equation}\label{R1ell}
\mathcal{R}_{1\ell}=\frac{2^{\ell+\frac{1}{2}}}{\beta^{\ell+\tfrac{3}{2}}}\Bigg[\Gamma(\ell+\tfrac{3}{2},\tfrac{\beta}{2}\rho_0^2)-\frac{\Gamma(\ell+\tfrac{5}{2},\tfrac{\beta}{2}\rho_0^2)}{(\ell+1)/2}+\frac{\Gamma(\ell+\tfrac{7}{2},\tfrac{\beta}{2}\rho_0^2)}{(\ell+1)^2}\Bigg].
\end{equation}

\textls[-30]{Being interested in the large-$\ell$ limit of $\mathcal{R}_{1\ell}$, we find in analogy with (\ref{M1ellInfinite}) and (\ref{P1ellInfinite}) that,}
\begin{align}\label{R1ellInfinite}
\mathcal{R}_{1(\ell\gg1)}&\approx\frac{2^{\ell+\frac{1}{2}}}{\beta^{\ell+\frac{3}{2}}}\left[ \Gamma(\ell+\tfrac{3}{2})-\frac{2\Gamma(\ell+\tfrac{5}{2})}{\ell+1}+\frac{\Gamma(\ell+\tfrac{7}{2})}{(\ell+1)^2}\right]
\approx\frac{2^{\ell+\frac{1}{2}}}{\beta^{\ell+\tfrac{3}{2}}}\frac{(\ell+\tfrac{7}{4})\Gamma(\ell+\frac{3}{2})}{(\ell+1)^2}.
\end{align}

The right-hand side of the last equality is used in Section~\ref{sec:V} to estimate the splitting of the relativistic energy levels for large orbitals $\ell$.

\subsection{Additional~Integrals}
For completeness, we display here the improper integrals involving two Kummer's functions with the integration boundaries $[0,\infty]$. Performing the change of variable, $x=\rho^2$, and~setting $b=\ell$ and $z=\beta/2$ in Equation\,(\ref{AppendixKummerIntegral1}), the~latter takes the following form,
\begin{equation}\label{AppendixKummerRhoIntegral1}
\int_{0}^{\infty}\rho^{2\ell+1}e^{-\frac{\beta}{2}\rho^2}\,_1F_1\left(-n,\ell+1;\frac{\beta}{2}\rho^2\right)\,_1F_1\left(-m,\ell+1;\frac{\beta}{2}\rho^2\right){\rm d}\rho
=\frac{n!2^\ell[\Gamma(\ell+1)]^2}{\beta^{\ell+1}\,\Gamma(n+\ell+1)}\delta_{nm}.
\end{equation}

Next, using identity (\ref{AppendixKummerLaguerre}), and~then performing again the change of variable, $x=\rho^2$, and~setting $c=\ell+\tfrac{1}{2}$, $z=\beta/2$, and~$b=\ell$, Equation\,(\ref{AppendixLaguerreIntegral2}) yields, 
\begin{multline}\label{AppendixKummerRhoIntegral2}
\!\!\!\!\!\!\!\!\!\!\int_{0}^{\infty}\!\!\rho^{2\ell}e^{-\frac{\beta}{2}\rho^2}\!\,_1F_1\left(\!\!-n,\ell+1;\frac{\beta}{2}\rho^2\right)\!\!\,_1F_1\left(\!\!-m,\ell+1;\frac{\beta}{2}\rho^2\right){\rm d}\rho\\
=\frac{2^{\ell-\frac{1}{2}}}{\beta^{\ell+\frac{1}{2}}}\frac{\Gamma\left(\ell+1\right)\Gamma\left(\ell+\frac{1}{2}\right)\Gamma\left(m+\frac{1}{2}\right)}{\Gamma\left(\frac{1}{2}\right)\Gamma(m+\ell+1)}\,_3F_2\left(-n,\ell+\frac{1}{2},\frac{1}{2},\ell+1,-m+\frac{1}{2};1\right).
\end{multline}

Finally, similar steps to those followed to get Equation\,(\ref{AppendixIntegralY}) allow us to prove that,
\begin{multline}\label{AppendixKummerRhoIntegral4}
\int_{0}^{\infty}\rho^{2\ell}e^{-\frac{\beta}{2}\rho^2-\frac{\rho}{\lambda}}\,_1F_1\left(-n,\ell+1;\frac{\beta}{2}\rho^2\right)\,_1F_1\left(-m,\ell+1;\frac{\beta}{2}\rho^2\right){\rm d}\rho\\
=\sum\limits_{k=0}^{\infty}\frac{(-1)^k}{k!\lambda^k}\frac{2^{\ell+\frac{k}{2}-\frac{1}{2}}\Gamma(\ell+\frac{k}{2}+\frac{1}{2})\Gamma(\ell+1)\Gamma(m+\frac{1}{2}-\frac{k}{2})}{\beta^{\ell+\frac{k}{2}+\frac{1}{2}}\Gamma(\frac{1}{2}-\frac{k}{2})\Gamma(m+\ell+1)}\\
\times\,_3F_2\left(-n,\ell+\frac{k}{2}+\frac{1}{2},\frac{k}{2}+\frac{1}{2};\ell+1,\frac{k}{2}+\frac{1}{2}-m;1\right).
\end{multline}

This last result, which, to~the best of our knowledge, has not been given previously in the literature on Kummer's functions either, is actually a generalization of the more familiar improper integral (\ref{AppendixKummerRhoIntegral2}) to the case of a Laplace transform involving an extra exponential function besides Kummer's~functions. 

\end{paracol}
\reftitle{References}

\end{document}